\documentstyle[aps,preprint]{revtex}
\begin{document}
\newcommand{\bra}[1]{\langle#1|}
\newcommand{\ket}[1]{|#1\rangle}
\newcommand{\braket}[2]{\langle #1 \vert #2 \rangle }
\newcommand{\be}[1]{\begin{equation}\label{eq:#1}}
\newcommand{\ee}{\end{equation}}
\newcommand{\req}[1]{(\ref{eq:#1})}
\newcommand{\bref}[6]{#2, {\sl #3}, {\bf #4}, #5, (#6)}
\parindent = 0pt
\draft
\tightenlines
\widetext
\title{Non-adiabatic transitions in multi-level systems}
\author{Michael Wilkinson$^1$ and Michael A. Morgan$^2$}
\address{\mbox{}$^1$
Department of Physics and Applied Physics,
John Anderson Building,
University of Strathclyde,
Glasgow, G4 0NG, UK.\\
\mbox{}$^2$
Department of Physics,
Seattle University,
Seattle, WA 98122, 
U.S.A.}
\date{\today}
\maketitle{ }
\begin{abstract}
In a quantum system with a smoothly and slowly varying Hamiltonian, which
approaches a constant operator at times $t\to \pm \infty$, the
transition probabilities between adiabatic states are exponentially
small. They are characterized by an exponent that depends on a phase
integral along a path around a set of
branch points connecting the energy level surfaces in complex time.
Only certain sequences of branch points contribute.
We propose that these sequences are determined by a topological rule
involving the Stokes lines attached to the branch points.
Our hypothesis is supported by theoretical arguments and
results of numerical experiments.
\vskip 1 cm
\noindent{University of Washington preprint NT@UW-98-26}
\end{abstract}

\tableofcontents

\section{Introduction}

Consider a quantum mechanical system with a Hamiltonian $\hat H(X)$
that depends analytically on a parameter $X$ and has a
discrete spectrum.  Suppose the parameter changes slowly and analytically with
time $t$, in such a way that asymptotically as $t\to \pm \infty$,
it remains constant. It is convenient to write $X=X(\tau )$ 
where $\tau = \epsilon t$. The {\it adiabatic parameter} $\epsilon$ 
is a small, real number which provides a dimensionless
measure of how slowly the Hamiltonian changes, due to the change in the
parameter $X$. The quantum adiabatic theorem states that if such 
a system is prepared in the $n^{\rm th}$ eigenstate of the
instantaneous Hamiltonian at any time, then in the limit $\epsilon\to 0$,
the system remains in the $n^{\rm th}$ eigenstate of the 
instantaneous Hamiltonian at later times [1,2].

Corrections to the adiabatic
theorem can be extremely small. If $X(\tau )$ approaches its limiting
values sufficiently rapidly as $\tau \to \pm \infty$, the probability
of making a transition to another eigenstate of the Hamiltonian vanishes
exponentially (faster than any power of $\epsilon$) as $\epsilon \to 0$.
The probability for making
a transition from the state $\ket {\phi_n(X(\tau_0))}$ to the state
$\ket {\phi_m(X(\tau_f))}$ is given asymptotically in $\epsilon$ by
$$P_{n\to m} \sim C_{nm}\exp [-2S_{nm}/\hbar]
\eqno(1.1)$$
where $C_{nm}$ and $S_{nm}$ are positive constants and the action
$S_{nm}$ is inversely proportional to $\epsilon$. Results of this
type have been proven rigorously for most cases of physical
interest in which the Hamiltonian is a $2\times 2$ [3,4] or
$3\times 3$ matrix [5,6], and for some special cases of multi-level
systems [7-10]. This paper will discuss how these
constants $S_{nm}$, $C_{nm}$ describing
 non-adiabatic transitions may be determined with some generality
for a discrete multi-level system. For definiteness, we will assume that
the Hamiltonian $\hat H$ is a symmetric
${\cal N}\times {\cal N}$ matrix, which is real for real $\tau$,
and analytic on a strip containing the real $\tau$ axis.
Then the Schwartz reflection principle
guarantees conjugacy of the matrix elements about the real time axis.

The special case of a two-level system has been discussed by
many authors. Zener [3] calculated an exact expression for the
transition probability for a two level model of the form
$$\hat H(X)={1\over 2}\pmatrix {X & \Delta \cr \Delta & -X \cr}
,\ \ \ X=A\tau
\ .\eqno(1.2)$$
This Landau-Zener system is of particular importance because for a
multi-level system, according to degenerate perturbation
theory, a near-degeneracy or \lq avoided crossing' of two levels can
be approximated by a two-state Hamiltonian of this form. The exact transition
probability for this system is
$$P_{LZ}=\exp (-\pi \Delta^2/2A\epsilon\hbar )\ .\eqno(1.3)$$
Note that the closest approach of the energy levels $E(t)$ is $\Delta $, so
that non-adiabatic transitions occur easily if the energy levels
approach each other closely.

Later Dykhne [4] showed that in the limit $\epsilon \to 0$ the
transition probability
for a general two level system is determined entirely by the analytic
continuation of the energy levels as a function of complex time. The
eigenvalues of the instantaneous Hamiltonian $E(\tau)$ form two branches
of a single complex function, which are connected at branch points 
occurring as complex conjugate pairs.
The action exponent
$S$ is  determined by integrating the energy along a path which leaves the
real axis and loops around a branch point so that it returns on the other
branch.  Dykhne showed that, in the limit $\epsilon\to 0$, the transition
probability becomes
$$P\sim \exp \biggl[ -{2\over \hbar}\;\Big|{\rm Im}\int_\gamma {dt\ E(t)} 
\Big|\,\biggr]\equiv \exp\bigl[-2S/\hbar\bigr] \eqno(1.4)$$
where $\gamma $ is the path illustrated in figure 1. The prefactor in
(1.1) is seen to be $C=1$. If there are more than one
pair of branch points, the one with the smallest
value of $S$ gives the dominant contribution to the transition probability.
For the model (1.2), $S=\pi \Delta^2/4\epsilon A$, so that the
Landau-Zener formula follows from this more general result.

In a multi-level system, it is natural to expect that the transition
probability is, in general, the product of several terms of the form of
(1.4), corresponding to several successive transitions between pairs
of levels.  The conclusions of this paper will confirm this
expectation. It might also be expected that the transition
probability will be determined by the combination of paths
around branch points that minimizes the total value of the
sum of integrals $S$ of the form appearing in (1.4). This expectation
is only partially correct: not all sequences of branch points
are allowed.  In this paper we propose a topological constraint, or rule, 
that the allowed combinations of branch points should satisfy. The value of
$S$ is determined by the minimal value of the sum of integrals of the form
(1.4) for paths consistent with the constraint.

An heuristic reason for the necessity of a rule that eliminates
certain sequences of branch points can be understood from an
example given by Hwang and Pechukas [5]. Consider a three level
system, with energy levels as plotted schematically in figure 2(a).
There are two avoided crossings where pairs of levels approach each
other very closely. In the neighborhood of these avoided crossings
the two nearly degenerate levels can be represented by a two-state
model similar to (1.2), and the transition probabilities between pairs of
adjacent levels are determined by the branch points near the real axis shown
in figure 2(b). The transitions from level $1$ to level $3$ occur through
Landau-Zener transitions at the successive avoided crossings, and the
transition probability is the product of two terms of the form of (1.4).
The transition from level $3$ to level $1$ cannot make use of the same
path, however, because the avoided crossing between levels $2$ and $1$
is encountered before the transition from level $3$ to level $2$ has
taken place. Instead, the transition occurs through a more distant branch 
point between levels $1$ and $3$ as shown in figure 2(b). The probability
for the $3\to 1$ transition is therefore expected to be much smaller. 
Therefore, a rule is required that eliminates the path from level $3$ to
level $1$ via the branch points closest to the real axis, but permits
this path for the reversed transition.

In section 2 a rule is proposed for determining which branch points
contribute to the transition probabilities. Our rule is motivated
by a discussion of the Stokes phenomenon given by Heading [11,12],
which assumes that evanescent waves are meaningful even in the
presence of larger contributions,
and that the amplitudes of evanescent waves change on the Stokes lines.
These assumptions must be justified by a reference to an asymptotic 
series in $\epsilon$ for the wavefunction, showing that the
errors in the dominant terms can be
reduced to below the magnitude of the subdominant (evanescent) terms,
except in the vicinity of the Stokes lines. This interpretation of the
asymptotic series for the wavefunction occurs in Stokes' analysis of the Airy
function [13], and many further examples are discussed in a book by Dingle
[14]. Works by Berry [15-17] have developed the theory further and discuss
physical applications, including quantum adiabatic theory for two-level 
systems.

The standard approaches for this interpretation of Stokes lines are valid
only for those Stokes lines associated with branch points having the smallest
action $S$, since these give the most divergent
contributions to the asymptotic series. All applications to date have
considered systems with only two states (in adiabatic theory) or two
channels (in WKB approaches to scattering problems), where only this one
branch point is relevant. These methods are bound to fail in multi-level
systems where many branch points are of necessity involved in a transition,
since the divergence of the series is dominated by the branch point with
the smallest action.  In section 3 we describe a formulation that allows all
of the Stokes lines to be given a similar interpretation.  By
projecting out the subspace corresponding to branch points with small
actions, divergences due to other branches are revealed.  The
method is based upon an exact iterative
\lq renormalization' of the Hamiltonian, adapted from a scheme
introduced by Berry [18]. Section 4 discusses the
limiting form of this series of renormalized Hamiltonian operators 
and its interpretation in terms of the Stokes phenomenon.
Section 5 contains some numerical illustrations of the results,
and section 6 summarizes the new contributions and considers further
development of the theory.

%
%
\section{A topological rule for determining the actions}

In this section we propose a topological rule for determining the
actions for non-adiabatic transitions motivated by a discussion
of the Stokes phenomenon.
The first three subsections review some ideas 
well known in the context of WKB theory.  We review these ideas
here to place them in the slightly different context of adiabatic theory
and to establish our notation.

\subsection{Adiabatic solutions of the Schr\"odinger equation}

It is possible to write down an asymptotic series for the solution of
the Schr\"odinger equation 
i$\hbar \, \partial_t \ket {\psi (t)}={\hat H(t)}\ket {\psi (t)}$ 
as an expansion in powers of $\epsilon$
$$\ket {\psi_n (t)}=\exp [-{{\rm i}\theta_n(t,t_0)/\hbar}]
\biggl\{ \ket {\phi_n(\tau )}+\epsilon \ket {\phi_n^{(1)}(\tau )}
+...+\epsilon^k\ket {\phi_n^{(k)}(\tau )}+...\biggr\}
\eqno(2.1)$$
where $\ket{\phi_n(\tau )}$ is an eigenvector of the
instantaneous Hamiltonian $\hat H(X(\tau))$, and $\theta_n(t,t_0)$
is the phase integral of the corresponding eigenvalue
($n^{\rm th}$ energy level) defined by
$$\theta_n(t,t_0)=\int_{t_0}^t dt'\ E_n(t')
={1\over \epsilon}\int_{\tau_0}^\tau d\tau \ E_n(\tau ) \ , \eqno(2.2)$$
with $t_0$ the phase reference.
The state in (2.1) is an approximate solution of the
time-dependent Schr\"odinger equation, assuming the system was 
prepared in the $n^{th}$ energy level in the remote past, that doesn't take 
into account non-adiabatic transitions.  Only the leading order terms in these
asymptotic expansions will be considered explicitly.  These will be termed the
{\it adiabatic solutions}
$$\ket {\psi_n (t)}=\exp[-{\rm i}\theta_n(t,t_0)/\hbar]\ket {\phi_n(t)}
\ .\eqno(2.3)$$
For complex $t$, the exponential factor need not have unit modulus.
As we move along a path in the complex $t$ plane, we can
compare the rate at which two solutions increase or decrease exponentially.
The solution for which the derivative of ${\rm Im}\,\theta_n$ is largest is
said to be {\it dominant} and the other solution {\it subdominant}.

Because the solutions are to be considered as functions
of a complex time variable, it is necessary to include
a few comments about how the eigenvalues and eigenvectors
are continued into the complex plane. Since for real $t$ the
matrix elements of the Hamiltonian are real, we define the inner
product of two eigenvectors without complex conjugation. We shall assume 
that the eigenvectors are normalized in the usual way, using
this inner product. As $t$ becomes
complex, the eigenvalues and eigenvectors also become complex,
but the Hamiltonian matrix remains symmetric and its eigenvectors orthogonal.
By choosing the eigenvectors real on the real axis, they then have no
gauge freedom. They may be expressed as analytic functions of $t$ at all 
points in the complex plane, except for a set of singular points.

Singular points of the eigenvalues and eigenvectors can arise
either because the matrix elements themselves have singularities,
or because of singularities in the mapping from matrix elements
to eigenvalues and eigenvectors.  This latter type of singularity
is associated with points where eigenvalues become degenerate
in the complex $t$ plane.
These singularities arising from degeneracies will be termed {\it branch
points}, because they connect two branches of the Riemann surface of
the eigenvalue function $E(t)$. The former type of singularities are of less
interest, because they are not universal in form, and because they
typically lie farther into the complex plane than the
branch points. (The reason for this will be discussed in
section 5.) Degenerate perturbation theory shows that in 
the neighborhood of two eigenvalues becoming degenerate, the
energy difference is the square root of the discriminant of the corresponding
quadratic equation.  At the singularity, the discriminant vanishes while,
generically, its time rate of change is non-zero. Thus generically,
singularities
due to eigenvalue degeneracies are square-root branch points.
In this paper we shall assume all relevant singularites are of this form.

For systems with a single parameter $X$, assumed here, a theorem of
Von Neumann and Wigner [19] states that, generically, degeneracies do not
occur  for real $t$.  Therefore the eigenvalues and eigenvectors can be
labeled with an index corresponding to the ordering of the (real)
eigenvalues on the real axis.  Usually it will be useful to consider the
eigenvalues and eigenvectors  to be single-valued functions of $t$.
This requires the introduction of branch cuts that restrict the domain
of definition of the function $E(t)$ to a single Riemann sheet. Typically
we will choose the branch cuts to be lines with constant ${\rm Re}\,t$ that
do not cross the real axis, as illustrated in figure 2(b).

It is necessary to establish how the eigenvalues and eigenvectors change
upon crossing a branch cut. Consider a branch point due to the degeneracy
$E_{n_1}=E_{n_2}$.  Since this is a square-root branch point,
the eigenvalues are simply exchanged as the branch cut is
traversed.  The eigenvectors are more involved.
If, starting from a point $t_+$ adjacent to the branch cut, the eigenvector
$\ket {\phi_{n_1}(t_+)}$ is followed anticlockwise around the branch point,
upon reaching the other side of the branch point the state
$\ket {\phi_{n_1}(t_-)}$ is equal to a multiple of the state
$\ket {\phi_{n_2}(t_+)}$. This is illustrated in figure 3(a).
Upon taking this multiple of the state $\ket {\phi_{n_2}(t_+)}$ anticlockwise
to $t_-$, it is transformed into $-\ket {\phi_{n_1}(t_+)}$, i.e., the
eigenvector
recurs with its sign changed after two circuits around the degeneracy.
This fact is easily verified in the special case of the Landau-Zener
model, (1.2).
Choose a branch cut crossing the real axis connecting the complex conjugate
branch points, and consider a circuit constructed
from the real axis and from a semicircle  at infinity, as illustrated in
figure 3(b).
Because the circuit can be shrunk to a circle enclosing the branch
point, and because the singularities of this model are generic,
the result is true in the general case.  Thus if we choose the phase of
the state $\ket {\phi_{n_2}}$ relative to $\ket {\phi_{n_1}}$ 
appropriately, we may write
$$\ket {\phi_{n_1}(t_-)}={\rm i}\ket {\phi_{n_2}(t_+)}$$
$$\ket {\phi_{n_2}(t_-)}={\rm i}\ket {\phi_{n_1}(t_+)} \eqno(2.4)$$
as a generic description of how the eigenvectors change when crossing a 
branch cut in the anticlockwise sense.

\subsection{Solutions in the neighborhood of a branch point}

This section considers the solution of the Schr\"odinger
equation in the vicinity of a branch point. The arguments are
an adaptation of those given by Heading [11,12] for the case of WKB
approximations. They are included here because the terminology
must be re-defined for the adiabatic problem, and because there
is a minor difference in the logic.

We assume that at $t^\ast $ the levels $E_{n_1}$ and $E_{n_2}$ become
degenerate. In the neighborhood of $t^\ast $
the difference between the two nearly degenerate levels behaves as
$\Delta E\equiv E_{n_2}-E_{n_1}\approx{\rm const.}(t-t^\ast )^{1/2}$, so that
the phase integral difference has a singularity of the form
$$\Delta \theta(t,t^\ast)=\int_{t^\ast }^{t}dt'\ \Delta E(t')
\approx {K_{t^\ast}\over {\epsilon}}(\tau-\tau^\ast )^{3/2}\eqno(2.5)$$
where $K_{t^\ast}$ is a complex constant characterizing
the branch point at $t^\ast$. Thus it is also necessary to introduce
branch cuts in the phase integrals $\theta_n(t,t^\ast)$,
which we take to be the same as for the energies.
On crossing a branch cut we must change the labeling of the energy
levels and phase integrals, although their values change
smoothly.

Some of the level curves (contours) of the phase integral that pass
through the branch point are very important in the formulation of the
theory. These are the Stokes and anti-Stokes lines defined by
$${\rm Re}\,\Delta \theta=0\ \ \ \ {\rm {Stokes\ lines}}$$
$${\rm Im}\,\Delta \theta=0\ \ \ \ {\rm {\textstyle {anti-Stokes\ lines}}}
\ .\eqno(2.6)$$
On the anti-Stokes lines the solutions (2.3) connected by the branch
point are co-dominant, and on the Stokes lines one solution can be
regarded as being maximally dominant over the other. The Stokes lines
associated with a typical branch point are sketched in figure 4.

The subdominant solution is meaningful only if it is larger than the
error term of the dominant series, which can itself be assumed to be
comparable to the smallest term of the asymptotic expansion (2.1).
On considering how the solution behaves near a branch point, it is
convenient to first consider the behavior of the solution on the
anti-Stokes lines.  This is because the adiabatic series corresponding
to the two levels which become degenerate at the branch point are co-dominant
on the anti-Stokes lines, and are therefore certainly both meaningful there.
If these solutions are
$\ket {\psi_1(t)}$, $\ket {\psi_2(t)}$, the solution at any point along
the anti-Stokes line (A1) in figure 4 may be written
$$\ket{\psi (t)}=\alpha_1\ket{\psi_1(t)}+\alpha_2\ket{\psi_2(t)}
\eqno(2.7)$$
with the multipliers $\alpha_1$, $\alpha_2$ approximately constant.
On the other two anti-Stokes lines the solution takes the same form,
but it need not have the same coefficients; for example, on line (A2)
the coefficients could be $(\alpha_1',\alpha_2')$. Since the
Schr\" odinger equation is linear, there must be a linear
relationship between the coefficients on the anti-Stokes lines:
$$\pmatrix{\alpha_1'\cr \alpha_2'\cr}
=\tilde M\pmatrix{\alpha_1\cr \alpha_2\cr }
\ .\eqno(2.8)$$
If $\ket{\psi_1(t)}$ is the dominant solution in the sector between
the lines (A1) and (A2), then the multiplier of this solution cannot
change, i.e., $\alpha_1'=\alpha_1$. Heading [11,12] has given a very
general argument, showing that the multiplier of the subdominant
solution is altered on crossing the Stokes line.  For the generic case,
where the singularity is given by (2.5) (triplets of Stokes lines attached
to a branch point), the argument follows closely that given by Heading
for the WKB approximation and the result is the same: the transition matrix 
from the anti-Stokes line (A1) to (A2) is
$$\tilde M=\pmatrix{1&0\cr -{\rm i}&1\cr }.\eqno(2.9)$$
In other  words, $-{\rm i}$ times the multiplier of the dominant solution
is added to that of the subdominant solution whenever a Stokes line is
crossed in the anticlockwise sense. This is {\it Heading's rule}. It assumes
a phase reference at the branch point.
Similar relations exist on crossing the other Stokes lines.
However, in writing down expressions for these relations it should be
remembered that, on crossing anti-Stokes lines, it is possible for the
dominance of the solutions
to switch.  For example, if $\ket{\psi_2(t)}$  is dominant on both (S2) and
(S3),
then dominance switches upon crossing (A1) and (A2) but not (A3).
The exact solution of the time-dependent Schr\" odinger equation
must be analytic and single valued
at the branch point, so equation (2.9) is obtained by
requiring that the multipliers $\alpha_i$
return to their original values when traced in a circuit around
the branch point, crossing three Stokes lines and the branch cut.
The branch cut is accounted for by noting that, according to (2.4),
upon crossing the cut the labels of the states are exchanged and the states
are multiplied by ${\rm i}$ (this circumstance differs from the case of
WKB theory, where the factors of ${\rm i}$ have a different origin).
We can introduce the matrix
$$\tilde T=\pmatrix {0&1\cr 1&0\cr}\eqno(2.10)$$
to account for the switching of dominance on the anti-Stokes lines.
The branch cut is accounted for by the matrix i$\tilde T$.
The effect of making a circuit around the branch point is then
described by the product i$\tilde T\tilde M\tilde T\tilde M\tilde T\tilde M$,
which is the identity matrix, verifying that
(2.9) is consistent with a single-valued wavefunction.

It is desirable to comment on the interpretation of (2.9).
The adiabatic wavefunctions are a poor approximation to
the exact solution at the branch point because they
are singular there. They are also a poor approximation far away from the
branch point because of the presence of other branch points and
singularities.  Furthermore, their interpretation
is ambiguous at all points off the anti-Stokes lines, because the subdominant
wavefunction may be smaller than the error of the dominant wavefunction. As
they stand then, the arguments above suffice only to discuss the behavior of
the wavefunction in a (deleted) neighborhood of the branch point, near the
anti-Stokes lines.

\subsection {Transition probabilities deduced from Heading's rule}

The results of section {\sl 2.2} are sufficient to enable the transition
probability due to a single branch point to be determined, provided that they
are supplemented by an additional assumption. The logic of the
argument presented in section {\sl 2.2} gives the form of the
wavefunction on two of the anti-Stokes lines, provided it is
known on the third (and even on the anti-Stokes lines it is
only a good approximation sufficiently far from the branch point).
The additional assumption is that the solutions may be extended
from the anti-Stokes lines as required and still
remain meaningful. This assumption requires the use of asymptotic
series approximations, which reduce the errors of the dominant
terms until the subdominant terms are meaningful.  Such approximations
are not discussed explicitly in Heading's work and must be
verified for specific applications.

It will be convenient to introduce the following notation for the
phase integral factor in the adiabatic solutions:
$$f_n(t,t_0)\equiv\exp[-{\rm i}\theta_n(t,t_0)/\hbar]
=\exp\biggl[-{{\rm i}\over{\hbar}}\int_{t_0}^tdt'\ E_n(t')\biggr]
\ .\eqno(2.11)$$
We assume that the system begins in the state
$\ket{\psi_1(t)}=f_1(t,t_0)\ket {\phi_1(t)}$ as $t\to -\infty$.
The phase reference $t_0$ is an arbitrary point on the real axis.
As $t\to \infty$, the system evolves to the state
$\ket{\psi_1(t)}+a\ket{\psi_2(t)}$, where $a$ is the transition amplitude
that we seek. In the domain in figure 4 bounded between (A1), (S1) and the
real  axis, the system is in the dominant state with respect to the branch
point in the upper-half plane. Heading's rule, given by (2.8) and 
(2.9), gives the evolution of the solution
around the branch point from (A1) to (A2) as
$$f_1(t,t^\ast)\ket {\phi_1(t)}\to f_1(t,t^\ast)\ket {\phi_1(t)}
-{\rm i}f_2(t,t^\ast)\ket {\phi_2(t)}
\ ,\eqno(2.12)$$
or in terms of the adiabatic solutions with phase reference $t_0$ as
$$f_1(t_0,t^\ast)\ket {\psi_1(t)}\to f_1(t_0,t^\ast)\ket {\psi_1(t)}
-{\rm i}f_2(t_0,t^\ast)\ket {\psi_2(t)}
\ .\eqno(2.13)$$
Assuming we may extend this solution down to time $t$ on the real axis, the
probability for the system to be found in state $\ket {\psi_2(t)}$ is
$$P_{1\to 2}\sim \vert f_2(t_0,t^\ast)f_1(t^\ast,t_0)\vert ^2=
\exp [-2\,{\rm Im}\,S_{1,2}/\hbar]
\eqno(2.14)$$
where
$$S_{1,2}\equiv
\int_{t_0}^{t^{\ast}}dt'\ \big[E_2(t')-E_1(t')\big]=-\int_\gamma
dt'E(t')
\eqno(2.15)$$
is the action evaluated around a path, analogous to that shown in figure 1,
encircling the branch point at $t^\ast$.
Its value is independent of the phase reference $t_0$.  Thus Dykhne's 
formula (1.4) has been deduced from Heading's rule.

Note that the branch point below the real axis does not make any
contribution to this transition probability, because the solution
$\ket {\psi_1}$ is subdominant in the lower half plane. To compute
the transition probability for transitions in the opposite direction,
the roles of the two conjugate branch points are reversed.

\subsection{The case of two branch points}

In this section we discuss the case in which a transition
occurs in two stages, first from the level $n_0$ to level $n_1$,
and then from level $n_1$ to level $n_2$. Consideration
of this case allows us to identify a rule for determining the
circumstances under which successive transitions may occur.

We examine a case in which adjacent pairs of levels become degenerate
at distinct branch points. If we are concerned with transitions from
lower to higher energies, branch points in the upper half plane are
relevant, because for paths below these branch points lower
lying levels are dominant. The Stokes lines (defined by (2.6))
of the two branch points can assume two possible arrangements 
as illustrated schematically in figures 5(a),(b).

First consider the arrangement in figure 5(a), with the initial
state being such that only the level with index $n_0$ is occupied
as $t\to -\infty$. Following the argument used in the last section,
we use Heading's rule to connect the solutions on the anti-Stokes
lines in the vicinity of the branch points, and assume that these
solutions may be extended down to the real time axis.  Since only
the initial state $\ket {\psi_{n_0}(t)}$ occupies the sector
between the lines (A1) and (S1) attached to the branch point
$(n_0,n_1)$, and since $\ket {\psi_{n_0}(t)}$ is the dominant
solution on (S1), Heading's rule gives the wavefunction on (A2) as
a multiple of $\ket {\psi_{n_0}(t)}-{\rm i}\ket {\psi_{n_1}(t)}$.
Thus the evolution of the solution around the $(n_0,n_1)$ branch
point  from (A1) to (A2) is given by
$$f_{n_0}(t,t_{n_0,n_1}^\ast)\ket {\phi_{n_0}(t)}\to
f_{n_0}(t,t_{n_0,n_1}^\ast)\ket {\phi_{n_0}(t)}
-{\rm i}f_{n_1}(t,t_{n_0,n_1}^\ast)\ket {\phi_{n_1}(t)}
\ .\eqno(2.16)$$
Next, since no Stokes lines are intervening, we may extend this
solution upward to the anti-Stokes line ${\rm (A1')}$ connected to
the $(n_1,n_2)$ branch point.  The sector between ${\rm (A1')}$
and ${\rm (S1')}$ is now occupied by the solution
$\ket{\psi_{n_1}(t)}$.  Since this is the dominant solution on ${\rm
(S1')}$,  Heading's rule gives the wavefunction on  ${\rm (A2')}$ as
a multiple of $\ket {\psi_{n_1}(t)}-{\rm i}\ket {\psi_{n_2}(t)}$.
The evolution of the solution around the $(n_1,n_2)$ branch point 
from ${\rm (A1')}$ to ${\rm (A2')}$ is given by
$$f_{n_1}(t,t_{n_1,n_2}^\ast)\ket {\phi_{n_1}(t)}\to
f_{n_1}(t,t_{n_1,n_2}^\ast)\ket {\phi_{n_1}(t)}
-{\rm i}f_{n_2}(t,t_{n_1,n_2}^\ast)\ket {\phi_{n_2}(t)}
\ .\eqno(2.17)$$
Combining these two evolutions, and expressing the result in terms of the
adiabatic solutions with a phase reference $t_0$ on the real axis, one obtains
$$\ket {\psi_{n_0}(t)}\to \ket {\psi_{n_0}(t)}
-{\rm i}f_{n_1}(t_0,t_{n_0,n_1}^\ast)f_{n_0}(t_{n_0,n_1}^\ast,t_0)
\ket{\psi_{n_1}(t)}$$
$$-f_{n_2}(t_0,t_{n_1,n_2}^\ast)f_{n_1}(t_{n_1,n_2}^\ast,t_{n_0,n_1}^\ast)
f_{n_0}(t_{n_0,n_1}^\ast,t_0)\ket{\psi_{n_2}(t)}\eqno(2.18)$$
for the evolution of the wavefunction from  ${\rm (A1)}$ to ${\rm
(A2')}$. Extending this solution down to time $t$ on the real 
axis, one finds the probability for making the
transition  from level $n_0$ to level $n_2$ is
$$P_{n_0\to n_2}\sim \vert
f_{n_2}(t_0,t_{n_1,n_2}^\ast)f_{n_1}(t_{n_1,n_2}^\ast,t_{n_0,n_1}^\ast)
f_{n_0}(t_{n_0,n_1}^\ast,t_0)\vert^2$$
$$=\vert f_{n_2}(t_0,t_{n_1,n_2}^\ast)f_{n_1}(t_{n_1,n_2}^\ast,t_0)
f_{n_1}(t_0,t_{n_0,n_1}^\ast)f_{n_0}(t_{n_0,n_1}^\ast,t_0)\vert^2$$
$$=\exp \bigl[-2\,{\rm Im}\, (S_{n_0,n_1}+S_{n_1,n_2})/\hbar\bigr]
\eqno(2.19)$$
where $S_{n_0,n_1}$ and $S_{n_1,n_2}$ are actions for paths
evaluated around the branch points located at times $t_{n_0,n_1}^\ast$ and
$t_{n_1,n_2}^\ast$, respectively. The transition probability is therefore
the product of two factors
of the Dykhne formula form, corresponding to two successive
transitions.

Now consider the arrangement in figure 5(b) with the same initial
state  $\ket {\psi_{n_0}(t)}$ occupied as $t\to -\infty$.  At the 
$(n_1,n_2)$ branch point, $\ket {\psi_{n_1}(t)}$ is dominant and  
$\ket {\psi_{n_2}(t)}$ is subdominant.  According to Heading's rule, 
a transition into the $n_2$ level requires that the subdominant 
solution be switched on in proportion to the dominant solution's 
multiplier. However, the solution $\ket{\psi_{n_1}(t)}$ has
multiplier zero in the entire region containing the relevant 
anti-Stokes lines attached to the $(n_1,n_2)$ branch point. 
It follows that no transition from $n_0$
to $n_2$ is possible using these two branch points.

There usually will be other branch points farther out in the 
complex plane connecting levels $n_0$ and $n_2$ directly. 
In both  of the cases discussed above, this
branch point would make a contribution to the transition 
probability of the form
$P_{n_0\to n_2}=P_{n_2\to n_0}=\exp[-2\,{\rm Im}\,S_{n_0,n_2}/\hbar]$.
In the case of figure 5(b), this would be the only contribution.
In the case of figure 5(a), it is the dominant contribution if
${\rm Im}\,S_{n_0,n_2}<{\rm Im}\,S_{n_0,n_1}+{\rm Im}\,S_{n_1,n_2}$, and
negligible otherwise.

\subsection{A criterion for selecting possible transition sequences}

We now propose a rule for determining the combinations
of actions that correspond to allowed transitions.
The rule is based upon Heading's local analysis
of the form of the solution in the vicinity of a branch
point, supplemented by the assumption that solutions may
be continued away from the branch points.

In the case of an upward transition from level $n_0$ to
$n_k$, a transition might be possible using a sequence
of branch points $t_{n_0,n_1}^\ast,\,
t_{n_1,n_2}^\ast,...,\,t_{n_{k-1},n_k}^\ast$
with transition probability
$P_{n_0\to n_k}=\exp[-2\,{\rm Im}\,(S_{n_0,n_1}+..+S_{n_{k-1},n_k})/\hbar]$.
This {\it transition sequence} is allowed only if the branch points
and their associated Stokes lines satisfy a topological criterion.
The transition $n_j\to n_{j+1}$ is mediated by the branch point
at $t_{n_j,n_{j+1}}^\ast$, and can occur only if the level $n_j$
is occupied at this branch point. This requires that the
branch point $t_{n_j,n_{j+1}}^\ast$ lies in the quarter plane 
above the real axis, and to the right of
the boundary formed by the Stokes line descending from
$t_{n_{j-1},n_j}^\ast$ to cross the real axis, and the
branch cut from the $t_{n_{j-1},n_j}^\ast$ branch point.

The rule is topological in character.
For downward transitions,
the relevant branch points are in the lower-half plane,
but because of reflection symmetry, the rule can be applied
in exactly the same way as for upward transitions.  The rule might
at first sight appear to have a degree of arbitrariness,
in that it refers to the positions of the branch cuts
as well as the Stokes lines. It may, however, be verified
that moving the branch cuts does not affect the predictions.
The reason is that when a branch cut from the $t_{n_0,n_1}^\ast$
branch point is moved past the $t_{n_1,n_2}^\ast$ branch point,
the labeling of the levels must be changed so that the
latter branch point now connects levels $n_0$ and $n_2$
rather than $n_1$ and $n_2$.

%
%
\section{Adiabatic renormalization with projections}

\subsection{Motivation for renormalizing the Hamiltonian}

The objective of this section is to explain how the interpretation
of the Stokes lines used in section 2, which led to the rule for the
selection of transition sequences, may be justified. This
requires the use of asymptotic series to reduce the error
of the dominant solutions as far as possible. It turns out that
this error can be reduced below the magnitude of the subdominant
solution everywhere, except in the vicinity of the Stokes lines.
This idea was introduced by Stokes [13] in a discussion of the
Airy function and amplified in a book by Dingle [14]. It has been
applied successfully in a variety of forms and to a variety of
physical problems by Berry [15-17]. To date, all of the applications
to differential equations have involved problems in which only two
equations are coupled together, e.g., two-state problems in adiabatic
theory, or one-dimensional semiclassical problems, involving two channels
(left and right propagating waves).

The approach used in these papers must be generalized
to cover the multi-level problem. The reason is that the
interpretation of the Stokes lines depends on studying the
divergence of an asymptotic series. This divergence is
usually determined by the branch point \lq closest' to
the real axis, in the sense of having the smallest
value of the imaginary part of the action, $\vert {\rm Im}\,S\vert$.
A means must be found to eliminate the divergence associated
with branch points close to the real axis in order that
the divergence associated with other branch points may
be revealed, allowing their Stokes lines to be interpreted.
This is achieved by \lq projecting out' a subspace of
the Hilbert space spanned by states having branch points
close to the real axis.

Various approaches were tried and found to be unsatisfactory.
The method we use is adapted from an approach introduced
by Berry [18]. Instead of constructing an asymptotic series
for the wavefunction, the objective is to construct a sequence
of \lq renormalized' Hamiltonians having the same dynamics as
the original problem and for which the adiabatic approximation is
successively more accurate. The $k^{\rm th}$ Hamiltonian
of this sequence has off-diagonal matrix elements of $O(\epsilon^k)$. If the
off-diagonal elements were to approach zero as $k\to \infty$, the adiabatic
approximation would be exact. In this case there would be no non-adiabatic
transitions. This implies that the prefactors
of the $O(\epsilon^k)$  terms must diverge as $k\to \infty$.
This divergence and its consequences are discussed in
section 4. In the remainder of this section we explain how the
sequence of renormalized Hamiltonians is constructed.
The approach follows that of Berry [18] quite closely, except that
a subspace of the spectrum is \lq projected out'
of the renormalization procedure, leaving its elements
$O(\epsilon^0)$.

\subsection{Renormalization excluding a subspace}

We consider a case where there is a branch point close to
the real axis in a subset of the spectrum characterized
by a set of state indices $\{P\}$. The projection operator
$\hat P(t)$ for the corresponding sub-space of the Hilbert
space is
$$\hat P(t)=\sum_{n\in P}\ket {\phi_n(t)}\bra {\phi_n(t)}
\ .\eqno(3.1)$$
The complementary set of state indices will be termed $\{Q\}$
and its corresponding projection operator designated $\hat Q=\hat I-\hat P$.
Provided the singularities of the matrix elements of
the Hamiltonian are sufficiently far from the real axis,
the singularities of the projection operator $\hat P$
are determined by the branch points where an eigenvalue
from $\{P\}$ becomes degenerate with one from $\{Q\}$.
The projection operator $\hat P(t)$ and the projected Hamiltonian
$\hat H_P=\hat P\hat H\hat P$ are analytic inside a strip $\Sigma_P$
symmetric  about the real axis and bounded by these conjugate branch
points.  This is true despite the fact that the eigenstates in (3.1)
have singularities closer to the real axis due to branch points
between states in $\{P\}$.

It will be useful to have available a representation
of the projection into the $\{P\}$ subspace 
explicitly constructed from analytic quantities. To this
end a set of states $\ket {\chi_n(t)}$, $n \in \{P\}$ will
be constructed having no singularities inside the strip
$\Sigma_P$ and satisfying
$$\braket {\chi_n(t)}{\chi_{n'}(t)}=\delta_{nn'}\ ,\ \ \
\hat P(t)\ket {\chi_n(t)}=\ket {\chi_n(t)}\ ,\ \ \
\lim_{t\to \pm \infty}
\ket {\chi_n(t)}
=\ket {\phi_n(\pm \infty)}
\ .\eqno(3.2)$$
This set of states could be constructed explicitly in one of several
ways. As an example, consider the  smooth interpolation
$${\textstyle{1\over 2}}(\ket {\phi_n(+\infty)}+\ket {\phi_n(-\infty)})
+{\textstyle{1\over 2}}(\ket {\phi_n(+\infty)}-\ket {\phi_n(-\infty)})
\,{\rm tanh}(\epsilon t)
\ .\eqno(3.3)$$
These state vectors could then be projected into the subspace by
multiplication by $\hat P$ and then  orthogonalized by a Gram-Schmidt
procedure. 

Consider the unitary operator
$$\hat U_0(t)=\sum_{n\in P}\ket {\chi_n(t)}\bra {\chi_n(-\infty)}
+\sum_{n \in Q} \ket {\phi_n(t)}\bra {\phi_n(-\infty)}\eqno(3.4)$$
which generates the states at $t$ from the states at $t=-\infty$. In order to
simplify notation, the states $\ket {\tilde\phi_n(t)}$ will be used to 
denote the states $\ket {\chi_n(t)}$ for $n \in \{P\}$ and $\ket {\phi_n(t)}$
for $n \in \{Q\}$ so that (3.4) is
$$\hat U_0(t)=\sum_n \ket {\tilde \phi_n(t)}\bra {\tilde \phi_n(-\infty)}
\ .\eqno(3.5)$$
Following Berry [18], we introduce a new representation
$\ket {\psi_1(t)}$ of the solution of the time-dependent
Schr\" odinger equation defined by
$$\ket {\psi(t)}=\hat U_0(t)\ket {\psi_1(t)}
\ .\eqno(3.6)$$
This renormalized wavefunction satisfies the Schr\" odinger equation
$${\rm i}\hbar \partial_t \ket {\psi_1}=\hat H_1\ket {\psi_1}
\eqno(3.7)$$
with the renormalized Hamiltonian given by
$$\hat H_1(t)=\hat U_0^\dagger(t)\hat H(t) \hat U_0(t)
-{\rm i}\hbar \hat U_0^\dagger(t)\partial_t \hat U_0(t)
\ .\eqno(3.8)$$
The matrix elements of $\hat H_1$ are conveniently evaluated
in the basis formed by the eigenvectors at $t\to -\infty$
$$H^{(1)}_{nm}\equiv
\bra { \phi_n(-\infty)}\hat H_1\ket { \phi_m(-\infty)}
=\bra {\tilde \phi_n(t)}\hat H\ket {\tilde \phi_m(t)}
-{\rm i}\hbar \braket {\tilde \phi_n(t)}{\partial_t \tilde \phi_m(t)}
\ .\eqno(3.9)$$
Three different cases arise. In the case where $n$ and $m$ are
both in $\{Q\}$
$$H^{(1)}_{nm}=\delta_{nm}E_n(t)-{\rm i}\hbar
{\bra {\phi_n(t)}\partial_t\hat H(t)\ket {\phi_m(t)}
\over{E_m(t)-E_n(t)}}(1-\delta_{nm})\ ,\ \ \ (n,m\in \{Q\})\eqno(3.10)$$
and for $n$, $m$ both in $\{P\}$
$$H^{(1)}_{nm}=\bra {\chi_n(t)}\hat H(t)\ket {\chi_m(t)}
-{\rm i}\hbar \braket {\chi_n(t)}{\partial_t \chi_m(t)}
\ ,\ \ \ (n,m\in \{P\})
\ .\eqno(3.11)$$
The case $n\in \{P\}$, $m\in \{Q\}$ requires some discussion.
The state $\ket {\partial_t \phi_m(t)}$ may be written
$$\ket {\partial_t\phi_m(t)}=\sum_{n\in \{P\}}a_n\ket {\chi_n(t)}
+\sum_{n\in \{Q\}}a_n\ket {\phi_n(t)}
\ .\eqno(3.12)$$
Differentiating the Schr\" odinger equation $(\hat H-E_m)\ket {\phi_m}=0$
with respect to time, then multiplying by $\bra {\chi_l}$ gives
the following linear equations for the $a_n$ with $n\in \{P\}$
$$\sum_{n\in \{P\}}D_{ln}a_n=b_l
\ ,\eqno(3.13)$$
where
$$D_{ln}\equiv E_m\delta_{ln}-\bra {\chi_l}\hat P\hat H\hat P\ket {\chi_n}
\ ,\ \ \ b_l\equiv\bra {\chi_l}\partial_t \hat H\ket {\phi_m}
\ .\eqno(3.14)$$
It is useful to define a matrix $\tilde G$ that is the inverse
of $\tilde D=\{D_{ln}\}$, and a corresponding operator $\hat G$
$$\hat G(t)\equiv \sum_{l,n\in \{P\}}\ket {\chi_l(t)}
G_{ln}(t)\bra {\chi_n(t)}=(E_m-\hat P\hat H\hat P)^{-1}
\ .\eqno(3.15)$$
Using the fact that $\tilde G=\{G_{ln}\}$ is the inverse
of $\tilde D$, one finds the solution of (3.13) is
$$a_l=\braket{\chi_l}{\partial_t \phi_m}=\sum_{n\in \{P\}}
G_{ln} \bra {\chi_n}\partial_t \hat H\ket {\phi_m}$$
$$=\sum_{n\in \{P\}}\bra {\chi_l}
(E_m-\hat P\hat H\hat P)^{-1}\ket {\chi_n}\bra {\chi_n}
\partial_t\hat H\ket {\phi_m}
\ .\eqno(3.16)$$
It follows that
$$H^{(1)}_{nm}=-{\rm i}\hbar \bra {\chi_n(t)}
(E_m-\hat P\hat H\hat P)^{-1}\hat P\partial_t \hat H\ket {\phi_m(t)}
\ ,\ \ \ (n\in \{P\},\ m\in \{Q\})
\ .\eqno(3.17)$$
In summary, the matrix elements of the renormalized Hamiltonian are given by the
expressions in (3.10), (3.11), and (3.17).  Notice that all of the off-diagonal elements
are $O(\epsilon)$, except those in the $n,m\in \{P\}$ block, which are $O(1)$, and that
$H^{(1)}_{nm}(t)$ is diagonal as $t\to \pm \infty$, because the
original Hamiltonian is constant in both limits.

\subsection{Iteration of the transformation}

The renormalization transformation defined in the last section can be
iterated by writing successive wavefunctions and Hamiltonian operators
as
$$\ket {\psi_k(t)}=\hat U_k(t)\ket {\psi_{k+1}(t)}\ , \ \ \
\hat H_{k+1}=\hat U_k^\dagger\hat H_k \hat U_k
-{\rm i}\hbar \hat U_k^\dagger
\partial_t \hat U_k
\ .\eqno(3.18)$$
Note that the transformation of the Hamiltonian is valid
for any choice of the unitary operator $\hat U_k$.
The operators $\hat U_k(t)$ will be defined by
$$\hat U_k(t)=\sum_n \ket {\tilde \phi^{(k)}_n(t)}
\bra {\tilde \phi^{(k)}_n(-\infty)}\eqno(3.19)$$
where
$$(\hat H_k-E^{(k)}_n)\ket {\phi_n^{(k)}}=0\ ,\ \ \ (n\in \{Q\})$$
$$\hat P_k\ket {\chi_n^{(k)}}=\ket {\chi_n^{(k)}}
\ ,\ \ \
\braket {\chi_n^{(k)}(t)}{\chi_{n'}^{(k)}(t)}=\delta_{nn'}
\ ,\ \ \ (n,n' \in \{P\})
\ .\eqno(3.20)$$
Here the new projection operator $\hat P_k$ is defined in analogy to (3.1)
using the new instantaneous eigenstates $\ket {\phi_n^{(k)}}$ for $n\in
\{P\}$.  The notation $\ket
{\tilde\phi_n^{(k)}(t)}$ is used to denote the states
$\ket {\chi_n^{(k)}(t)}$ for $n \in \{P\}$ and $\ket {\phi_n^{(k)}(t)}$
for $n \in \{Q\}$.  
Since both $H^{(1)}_{nm}(\pm \infty)$ are diagonal,
we have $\ket {\tilde \phi^{(1)}_n(\pm \infty)}=\ket {\phi_n(-\infty)}$.
The $\ket {\chi_n^{(k)}(t)}$ are chosen so that they are analytic in the
strip $\Sigma _{P_k}$ and close to the $\ket {\phi_n(-\infty)}$ as
$t\to\pm\infty$.  As in the last section for the $k=0$ case, these
states can be generated by acting on linear combinations of the  $\ket
{\phi_n(\pm\infty)}$ with the projection operator
$\hat P_k(t)$, then using a Gram-Schmidt procedure to create
orthonormalized states.
It is still most convenient to evaluate
matrix elements of the Hamiltonian $\hat H_k$ with respect to the states
$\ket {\phi_n(-\infty)}$. Generalizing the equations in section {\sl 3.2},
the matrix elements $H^{(k+1)}_{nm}\equiv\bra
{\phi_n(-\infty)}\hat H_{k+1}(t)\ket {\phi_m(-\infty)}$ are given by
$$
H^{(k+1)}_{nm}=\delta_{nm}E_n^{(k)}(t)-{\rm i}\hbar
{\bra {\phi_n^{(k)}(t)}\partial_t\hat H_k(t)\ket {\phi_m^{(k)}(t)}
\over{E^{(k)}_m(t)-E_n^{(k)}(t)}}(1-\delta_{nm}),
\ \  (n,m\in \{Q\})
\eqno(3.21a)$$
$$=\bra {\chi_n^{(k)}(t)}\hat H_k(t)\ket
{\chi_m^{(k)}(t)}-{\rm i}\hbar \braket {\chi^{(k)}_n(t)}
{\partial_t\chi_m^{(k)}(t)},
\ \ \ \ \ \ \ \ \ (n,m\in \{P\})
\eqno(3.21b)$$
$$=-{\rm i}\hbar \bra {\chi_n^{(k)}(t)}
(E_m^{(k)}-\hat P_k\hat H_k\hat P_k)^{-1}\hat P_k\partial_t
\hat H_k\ket {\phi_m^{(k)}(t)}. \ \ \ \ \ \ (n\in\{P\},m\in\{Q\})
\eqno(3.21c)$$
After $k$ stages of iteration all matrix elements are $O(\epsilon^k)$, except
the diagonal elements and all of the elements of the $PP$ block, which remain
$O(1)$. Each Hamiltonian is an exact representation of the dynamics of the
original problem, and the transition amplitudes could, in principle,
be obtained by integrating the Schr\" odinger equation
with any Hamiltonian in the sequence.  At each stage the 
$H^{(k)}_{nm}(\pm \infty)$ are diagonal and  hence $\ket
{\tilde\phi^{(k)}_n(\pm \infty)}=\ket {\phi_n(-\infty)}$. 
The amplitude for transition from state $n$ to state $m$ is therefore given
by $\braket {\phi_m(\infty)}{\psi_n(\infty)}=\braket
{\phi_m(-\infty)}{\psi_{k,n}(\infty)}$, where $\ket {\psi_n(t)}$ and
$\ket {\psi_{k,n}(t)}$ are the wavefunctions obtained by
propagating the initial state $\ket {\phi_n(-\infty)}$
with Hamiltonians $\hat H$ and $\hat H_k$, respectively. 
If the $O(\epsilon^k)$ matrix elements vanished as $k\to \infty$ 
there would be no non-adiabatic transitions. Therefore, the sequence 
of renormalized Hamiltonians is expected to
have the typical behavior for terms of an asymptotic series, in that, 
although the matrix elements decrease for small
$k$, at sufficiently large $k$ they diverge because
of a faster than exponential growth of the prefactors.
The value of $k$ for which the largest of the small
matrix elements has smallest magnitude will be denoted by
$k^\ast(\epsilon)$.
This divergence will be examined in greater detail
in section 4.

For small $\epsilon$ and for large values of $k$ which are
not too large, the off-diagonal matrix
elements outside the $PP$ block are very small.  Thus, for states in $\{Q\}$, 
the approximations 
$$\ket {\phi_n^{(k)}(t)}\sim \ket {\phi_n(-\infty)}\ , 
\ \ \ \ \ E_n^{(k)}(t)\sim E_n(t)\ ,\ \ 
(\epsilon \ll 1,\ \ 1\ll k\le k^\ast,\ \ n\in \{Q\})
\ \eqno(3.22a)$$
are valid, while for states in $\{P\}$ we are at liberty to choose 
$$\ket {\chi_n^{(k)}(t)}\sim \ket {\phi_n(-\infty)}\ ,\ \
(\epsilon \ll 1,\ \ 1\ll k\le k^\ast,\ \ n\in \{P\})
\ .\eqno(3.22b)$$
From this, along with (3.11) and (3.21b), it follows that for small $\epsilon$
$$\bra {\chi_n^{(k)}(t)}\hat H_k(t)\ket {\chi_m^{(k)}(t)}
\sim \bra {\chi_n(t)}\hat H(t)\ket {\chi_m(t)}\ ,\ \ \ (n,m\in \{P\})
\ .\eqno(3.23)$$
When these approximations are valid, equations (3.21) simplify considerably.
The basis states are approximately time independent, so the derivatives
with respect to the Hamiltonian become derivatives with respect to
matrix elements
$$H_{nm}^{(k+1)}
\sim\delta_{nm}E_n(t)-{\rm i}\hbar{\partial_t[H_{nm}^{(k)}(t)]
\over{E_m(t)-E_n(t)}}(1-\delta_{nm})\ ,\ \ \ \ \
(n,m\in\{Q\})
\eqno(3.24a)$$
$$\sim H^{(k)}_{nm}(t)\ ,
\ \ \ \ \ \ \ \ \ \ \ \ \ \ \ \ \ \ \ \ \ \ \ \ \ \ \ \ \ \ \ \ \ \ \ \ \ \
\ \ \ \ \ \ \ \
(n,m\in\{P\})
\eqno(3.24b)$$
$$\sim -{\rm i}\hbar \sum_{l\in \{P\}}G_{nl}\,
\partial_t [H^{(k)}_{lm}]\ ,
\ \ \ \ \ \ \ \ \ \ \ \ \ \ \ \ \ \ \ (n\in \{P\},\ m\in \{Q\})\ .
\eqno(3.24c)$$
To summarize: equations (3.21) describe an exact
renormalization of the Hamiltonian, which for
sufficiently large (but not too large) $k$, has the effect of reducing
the magnitude of all of the off-diagonal elements
outside the $PP$ block. Equations (3.24) are an approximate
implementation of this iteration valid for sufficiently
small $\epsilon$.

%
%
\section{Interpretation of the asymptotic series}

All of the renormalized Hamiltonians $\hat H_k$ describe the
same dynamics. In particular
$\vert \braket {\phi_m(-\infty)}{\psi_{k,n}(\infty)}\vert^2$,
where $\ket{\psi_{k,n}(t)}$ is the wavefunction propagated
under the Hamiltonian $\hat H_k$ from the initial state
$\ket {\phi_n(-\infty)}$, is the transition probability from state
$n$ to state $m$ and is independent of $k$. Now the viewpoint of
section 2, which associates transitions between pairs of levels
with Stokes lines, will be confirmed here by showing that when
$k$ is suitably large, the Hamiltonian $\hat H_k(t)$ on the real
axis is greatest at crossings of Stokes lines [17]. Section
{\sl 4.1} examines the case where there is no projected subspace: 
a direct generalization of the result for two level
systems. Section {\sl 4.2} considers the extension to the case
where there is a projected subspace, showing that the Stokes
lines can still have the same significance for branch points 
not close to the real axis.
Section {\sl 4.3} describes how these results can be
used to calculate a transition probability involving
a sequence of branch points.

\subsection{Form of the Hamiltonian at large order, excluding
projected subspace}

This section discusses the form of the matrix elements of the
renormalized Hamiltonian in the limit where $\epsilon \ll 1$
and $1\ll k \le k^\ast$. Equation (3.24a) which determines the
evolution of off-diagonal matrix elements in the $QQ$ block, is an
iteration of the form
$$H_{k+1}(t)=-{{\rm i}\hbar\over {\Delta(t)}}\partial_t H_k(t)
\ \eqno(4.1)$$
where $H_1(t)$ and $\Delta(t)$ are given functions. 
From observations on asymptotic series discussed by Dingle [14] it is 
well known that for large $k$, solutions of (4.1) may be obtained in the 
form
$$H_k(t)\sim (-1)^k A\, \Gamma (k+\gamma ) [F(t)]^{-(k+\gamma)}
\ \eqno(4.2)$$
where $A$ and $\gamma $ are constants, and $\Gamma (x)$ is the
Gamma function satisfying $\Gamma (x+1)=x\Gamma(x)\equiv x!$.
Substitution of (4.2) into (4.1) shows that the dimensionless scalar 
function $F(t)$ is
$$F(t)={{\rm i}\over \hbar}\int_{t^{\ast}}^t dt'\ \Delta (t')\ ,\eqno(4.3)$$
and the lower limit $t^{\ast}$ will sometimes be shown
explicitly as $F(t,t^{\ast})$.
The possible values of $t^{\ast}$ are determined by realizing that
an acceptable solution of (4.1) should be regular at any
point for which $\Delta^{-1} (t)$ and $H_1(t)$ are regular.
The solution (4.2) is clearly divergent for $k+\gamma >0$
when $t\to t^{\ast}$. This observation implies that $t^{\ast}$ must
correspond to a singular point, where at least one of
$\Delta^{-1} (t)$ and $H_1(t)$ has non-analytic behavior.  In the context 
of adiabatic theory $t^{\ast}$ corresponds to a branch point 
singularity.

In general, the solution at time $t$ is dominated by the approximate
solution (4.2) attached to a singular point $t^{\ast}$ for which the
modulus  of $F(t,t^{\ast})$ is smallest.  Since equation (4.1) is linear,
a superposition of co-dominant solutions of the form (4.2) may be
necessary. In the case where the functions $H_1(t)$ and $\Delta^{-1} (t)$
are analytic and real valued on the real axis, the singular points
occur as complex conjugate pairs.  In this situation the solution on the
real axis is a superposition of solutions with conjugate
singularities.

In our application where $\Delta = (E_m-E_n)$, the real part of $F$ is 
constant along the real axis.  Thus, on the real axis, the magnitude of 
the solution (4.2) is largest at
the point where the Stokes line, defined by
${\rm Im}\,F(t,t^{\ast})=0$, crosses.  For large $k$ the magnitude
decays rapidly, with an  approximately gaussian form on either side
of this point. When the two conjugate solutions are combined, the
result is a real function with oscillations within an approximately
gaussian envelope.

The constants $A$ and $\gamma $ can be determined - although
they are not required for application of the theory.
Both constants are obtained by considering the behavior
of the functions in the neighborhood of the singularity.
In the application of these results to adiabatic theory, 
equation (3.10) requires that the function $H_1(t)$ plays the 
role of $\bra {\phi_n(t)}\partial_t\hat
H\ket{\phi_m(t)}/{(E_m(t)-E_n(t))}$,  which scales as 
$c(t-t^\ast)^{-1}$ in the vicinity
of a branch point singularity at $t^\ast$. In the vicinity 
of the branch point,
the form of the successive $H_k(t)$ is dominated
by the components of $H_1(t)$ and $\Delta^{-1} (t)$ which are
most strongly divergent as $t\to t^\ast$. It is therefore
appropriate to take $H_1(t)=c(t-t^\ast)^{-1}$ and
$\Delta (t)=\alpha (t-t^\ast)^{1/2}$. This leads by direct
application of (4.1) to a general form for the $H_k(t)$
valid in the neighborhood of $t^\ast$
$$H_k(t)=C_k(t-t^\ast)^{-(3k-1)/2}\ ,\ \ \ C_{k+1}=-{3k-1\over 2\alpha}C_k
\ .\eqno(4.4)$$
The iteration for the coefficients $C_k$ has a solution
$$C_k=\biggl({-3\over
{2\alpha}}\biggr)^{k-1}\ {\Gamma(k-{\textstyle{1\over 3}})
\over{\Gamma({\textstyle{2\over 3}})}}c
\ .\eqno(4.5)$$
Comparison of this result with (4.2) shows that they are
consistent in the limit $k\to \infty$, and that
$\gamma=-{1\over 3}$ and
$A=-({{2\alpha}\over 3})^{2\over 3}c/\Gamma({2\over 3})$.

Applying these results to the renormalization of the $QQ$
block of the Hamiltonian (equation (3.24a)), 
shows that for large $k$ the off-diagonal matrix elements
are largest for the element
$H^{(k)}_{nm}$ for which $\vert F \vert$ is
smallest. On the real axis $F={\rm i} S_{nm}/\hbar$, so this
corresponds to the matrix element with the least 
$\vert {\rm Im}\,S_{nm}\vert$.
Furthermore, each of these matrix elements is largest in the vicinity of the
point where the Stokes  line intersects the real axis. The most significant
process in the  dynamics of $\hat H_k$ is therefore a 
transition between levels
$n$ and $m$  localized in time at the point where the Stokes line crosses the
real axis. The use of the renormalization scheme therefore confirms the
interpretation that the transition occurs at the Stokes line. The results are
consistent with the conclusions of section 2, and we  will assume that the
transition probability is given by (2.14) and (2.15).

\subsection{Form of the Hamiltonian at large order, including
projected subspace}

In the case where there is a sub-space projected out
in the manner discussed in section 3, the equation for the
iteration of the off-diagonal matrix elements in the $QQ$ block is
exactly of the form considered in section {\sl 4.1} above. The
case of matrix elements in the $PQ$ and $QP$ blocks requires
a different treatment. The equation for iteration of the
these elements of the Hamiltonian is of the form (3.24c),
which we write symbolically as
$$\tilde H_{k+1}(t)=-{\rm i}\hbar\,\tilde G(t)\,\partial_t \tilde H_k(t)
\eqno(4.6)$$
where $\tilde H_k$ is a column vector of dimension ${\cal N}_P$,
with elements $H^{(k)}_{nm}$, and
$\tilde G=\{G_{nl}\}
=\{\bra {\chi_{n}}(E_m-\hat P\hat H\hat P)^{-1}\ket {\chi_l}\}$, a matrix
of dimension ${\cal N}_P\times {\cal N}_P$.  Note that the index $m$ is
fixed and not shown in (4.6).  Following (4.2), it might be anticipated
that a solution of (4.6) is of the form
$$\tilde H_k=(-1)^k\,\Gamma (k+\gamma)
\bigl[\tilde M(t)\bigr]^{-(k+\gamma)}\tilde A
\ ,\eqno(4.7)$$
where $\tilde M(t)$ is a matrix of
dimension ${\cal N}_P\times {\cal N}_P$ and $\tilde A$ a 
constant column vector of dimension
${\cal N}_P$.  We proceed to show that (4.7) is the solution 
to (4.6) to leading order in $k^{-1}$. Let
$\tilde M(t)=\tilde X^{-1}(t)\tilde F(t)\tilde X(t)$, where 
$\tilde F(t)$ is diagonal. 
Then the time derivative may be simplified using the following approximation
valid to leading order in $k^{-1}$:
$$\partial_t \bigl[\tilde M(t)\bigr]^{-k}=-k\tilde X^{-1}(t)
\partial_t \tilde F(t)\bigl[\tilde F(t)\bigr]^{-(k+1)}
\tilde X(t)+O(1)
\ .\eqno(4.8)$$
%
%
%
%
Using this approximation one finds that (4.7) is the solution 
to (4.6) to leading order
in $k^{-1}$ so long as 
$-{\rm i}\hbar \tilde X \tilde G \tilde X^{-1}\partial_t \tilde F$ 
is the identity matrix. It follows that 
$\tilde X \tilde G \tilde X^{-1}$ must be
diagonal since $\tilde F$ is assumed to be diagonal. 
The matrix $\tilde X(t)$ is
therefore the similarity transformation that diagonalizes $\tilde G(t)$,
and the matrix $\tilde F(t)$ and its elements $F_n(t)$ are
$$\tilde F(t)={{\rm i}\over \hbar}\,\int_{t^{\ast}}^t dt'\, \tilde X(t')\, 
\tilde G(t')^{-1}\,\tilde X^{-1}(t')$$
$$F_n(t)={{\rm i}\over \hbar}\,\int_{t^{\ast}}^t dt'\
\bigl[ E_m(t')-E_n(t') \bigr]
\ .\eqno(4.9)$$
Substitution into (4.7) now gives the required approximation.
These results are a natural generalization of the results
for the simpler case described by the iteration (4.1).
A similar argument to that given in section {\sl 4.1} forces
$t^{\ast}$ to be the position of a singularity, where a pair
of eigenvalues, one each from the $P$ and $Q$ subspaces,
become degenerate.

It is worth noting that these expressions can be simplified
by making use of freedom available in the specification
of the states $\ket {\chi_n(t)}$. The states
$\ket {\chi_n(t)}$ may be chosen to coincide with the eigenstates
$\ket {\phi_n(t)}$ at isolated positions in the complex
time plane, so that the similarity transformation $\tilde X(t)$
is the identity at that point. Specifically, these states
may be chosen to coincide at the point where a Stokes line
crosses the real axis.
The comments in section {\sl 4.1}, indicating
how the renormalized Hamiltonian is concentrated on Stokes
lines, can then be applied equally well to this case: from (4.7) and (4.9)
it can be seen that the dominant term in the high-order
renormalized Hamiltonian comes from the branch point
with the smallest value of $\vert {\rm Im}\, S_{nm}\vert$, where at least
one of $n$, $m$ is in the $Q$ subspace. If $\tilde X(t)$
is the identity  where the corresponding Stokes line crosses the
real axis, the largest matrix elements determine
the transition between states $n$ and $m$.

\subsection{Interpretation of the renormalized Hamiltonian}

As an example of how these considerations can be used to
demonstrate the validity of the topological rule, consider
their application to the case of two successive transitions
$n_0\to n_1$ followed by $n_1\to n_2$. It will be assumed
that the $(n_0,n_1)$ branch point is closest to the real axis 
and that the $(n_1,n_2)$ branch point is next closest.
The subspace $P$ is then chosen to be that spanned by
levels $n_0,\,n_1$. The states $\ket {\chi_{n_0}(t)}$
and $\ket {\chi_{n_1}(t)}$ are chosen to correspond
to the eigenstates $\ket {\phi_{n_0}(t)}$ and
$\ket {\phi_{n_1}(t)}$ at the point $t_S$ where the
Stokes line from the $(n_1,n_2)$ branch point crosses the
real axis. When the order $k$ is sufficiently large, the
matrix elements that determine transitions from the $P$ subspace
are largest for the transition between levels $n_1$ and
$n_2$, and are concentrated at $t_S$. The general
arguments given in section 2 indicate that this transition
probability is
$P_{n_1\to n_2}\sim \exp[-2\,{\rm Im}\,S_{n_1,n_2}/\hbar]$.

The probability
for making a transition from state $n_0$ to $n_2$ via the
intermediate state $n_1$ therefore depends on the dynamics
within the $PP$ sub-block at times earlier than $t_S$.
The matrix elements within this sub-block have been left
approximately unchanged by iteration of the renormalization
procedure. Any suitable procedure can be used to calculate the
probability for the transition $n_0\to n_1$ occurring
before the time $t_S$. In the limit $\epsilon \to 0$, the most
convenient procedure is of course to use adiabatic theory
that predicts the transition will occur on the Stokes
line attached to the $(n_0,n_1)$ branch point with the probability
$P_{n_0\to n_1}\sim \exp[-2\,{\rm Im}\,S_{n_0,n_1}/\hbar]$.
The overall transition probability for the $n_0\to n_2$ transition
going through the intermediate state $n_1$ is then
$P_{n_0\to n_2}=P_{n_0\to n_1}P_{n_1\to n_2}$, if the branch point
$(n_1,n_2)$ lies to the right of the Stokes line from
the $(n_0,n_1)$ branch point. In the other case, it is determined
by other branch points.

%
%
\section{Numerical illustration}

The scenario described above was tested numerically on a model
Hamiltonian $\hat H(X(\tau))$ of the following form:
$$\hat H=\cos (X(\tau ))\hat H_1+\sin (X(\tau ))\hat H_2\eqno(5.1)$$
$$X(\tau )=\alpha \tanh ({\tau / \alpha} )\eqno(5.2)$$
where $\tau = \epsilon t$, and $\hat H_1$ and $\hat H_2$ are
two independent random square matrices of dimension $\cal N$,
drawn from the gaussian orthogonal ensemble (GOE). The GOE
ensemble consists of real, symmetric matrices with independently
gaussian-distributed elements with mean and
variance given by [20,21]
$$\langle H_{ij}\rangle =0,\ \ \ \langle H_{ij}^2\rangle =(1+\delta_{ij})
\ .\eqno(5.3)$$
The choice of Hamiltonian is arbitrary, so long as its spectrum is
nondegenerate for all real $\tau$, and its eigenvalues and eigenstates
become time independent asymptotically as $\tau\to\pm\infty$. The
function $X(\tau )$ was chosen to fulfill the latter requirement.
The model (5.1) was used because, for ${\cal N}\gg 1$,
its spectral properties are representative
of those of generic time-reversal invariant multi-level systems [22].
The distance of the branch point singularities from the real
axis scales as ${\cal N}^{-1/2}$, whereas the singularities
of the matrix elements in the complex $\tau $ plane have
a distribution of positions independent of the
matrix dimension ${\cal N}$. It follows that in typical
physical applications branch points will lie closer to the real
axis than other singularities.

Numerical calculations were performed to determine the \lq exact' or
\lq empirical' transition  probabilities $P_{n \to m}$ for $100$
Hamiltonians of the form  (5.1), all with dimension ${\cal N}=6$.
In each case all ${\cal N}({\cal N}-1)=30$ transition
probabilities were computed.  The scale factor $\alpha= 2$ was assumed
in all calculations.  The probabilities were obtained by numerical time
integration of the Schr\"odinger equation, using a standard fourth order
Runge-Kutta algorithm with  arithmetic accurate to $14$ decimal places.
Rather than comparing  the transition probabilities themselves, we
compared (the imaginary part of) the actions times the adiabatic
parameter $\epsilon$ in units of Planck's constant:
$$\lambda_{nm}\equiv {{\vert {\rm Im}\,{S_{nm}\vert}\over \hbar}\>
\epsilon}=-{1\over 2}\log(P_{n\to m})\>\epsilon
\ .\eqno(5.4)$$
The integration  was performed between $\tau_i=-25$ and $\tau_f=+25$.
We checked that the transition probabilities are insensitive to
further increasing the range of $\tau$. These calculations
were done for a large set of adiabatic parameters between
$\epsilon = 0.001$ and $\epsilon = 0.5$. We observed that in most
cases the actions converged to a limit as $\epsilon \to 0$. In other
cases, we estimated the limit by extrapolation with a polynomial. In
many cases numerical roundoff error would make the results unreliable
for small $\epsilon$. In these cases we assumed the best value for the 
action corresponded to the smallest value of $\epsilon$ for which
roundoff errors were not significant. Typically, roundoff errors were
significant when the calculated result obeyed $P_{n\to m}\le 10^{-23}$.

The adiabatic calculations were performed in the following way. First, the
approximate locations of the branch points $\tau_{i,j}^{\ast}$ were  
determined by a search for near-degeneracies on a grid in the complex
$\tau$ plane. The rectangular region bounded by $-6\le {\rm Re}\> \tau
\le 6$  and ${0\le {\rm Im}\> \tau }\le 2$ was usually found to include
all the relevant branch points. Next, the locations of these branch point
candidates were refined by a version of the Newton-Raphson method 
adapted to finding square root branch points of the form
${E_i-E_j} \propto {\sqrt {\tau - \tau_{i,j}^{\ast}}}$.
The Stokes and anti-Stokes lines were plotted for several
realizations of the random Hamiltonian, and the allowed
transition sequences were determined. The Stokes and anti-Stokes lines
attached to the branch points were traced by evaluating a sequence of
short steps $\delta t$ along their lengths. If the Stokes line emerging
from a branch point between levels $i$ and $j$ was found to pass
through the point
$\tau_k$, the next point was obtained from $\tau_{k+1}=\tau_k+\delta \tau_k$
where from (2.6) 
$\delta \tau_k= \varepsilon {\rm i}/[E_i(\tau_k)-E_j(\tau_k)]$, and
$\varepsilon $ is a small real number. Increments of the
anti-Stokes lines were determined by an analogous approach.
Finally, the \lq adiabatic' or
\lq theoretical' branch point actions were computed numerically
according to
$$\lambda_{i,j}={1\over \hbar}\>\Big|{\rm Im}\int_{({\rm
Re}\,\tau_{i,j}^{\ast},0)}^{\tau_{i,j}^{\ast}}d\tau\>\big(E_i(\tau) -
E_j(\tau)\big)\>\Big|
\ .\eqno(5.5)$$
\par
The theoretical
actions were in excellent agreement with those determined
empirically from (5.4). We found that in every case 
the allowed transition sequence could have been determined
by a simple empirical rule, namely, that the real parts
of the branch points should be in ascending order.
Accordingly, the branch point data were ordered with respect to
${\rm Re}\, \tau_{i,j}^{\ast}$. From this ordered table of branch
point locations and branch point  actions, all possible transition
sequences from the initial state $n$ to the final state $m$, in which the
state indices increased (or decreased) monotonically, were
considered.  According to the rule given in section {\sl 2.5}, the
overall action $\lambda_{nm}$ was taken to be the least sum
of (the imaginary part of) the branch point actions over these
possible transition sequences.  
These calculations of the overall  \lq theoretical' actions were
automated.

Examples of the calculations are presented in the tables and figures.
Figures $6$ and $7$ show plots of the energy eigenvalues for real time for
two examples of the $100$ sample Hamiltonians. The branch point data
corresponding to these two cases are listed in tables $1$ and $2$. Observe 
that the avoided crossings in these figures correspond to small values of
the branch point actions in the tables.  Figures $8$ and $9$ show the
Stokes lines that cross the real axis for all branch points involved in
some transition sequence. The  \lq theoretical' transition sequences and
actions  derived from these data are shown in tables $3$ and $4$. The \lq
empirical'  results obtained using (5.4) are shown there for comparison.
Note that the fractional difference between the empirical and theoretical
results based on our adiabatic theory is typically around
$1\%$ or smaller, although occasionally as high as $2\%$.  The average
fractional difference between the empirical and theoretical actions over
all $3{,}000$ data points was $\sim 1.6\%$.  The data are consistent  with
the hypothesis that the transition probabilities are given by (1.1) with
$C=1$, and the action $S_{nm}$ given by the topological rule of
section {\sl 2.5}.

%
%
\section{Concluding remarks}

This paper makes three novel contributions to understanding the
behavior of the transition probabilities for multi-level systems.
First, we have suggested a general rule for determining the
combinations of branch points that give allowed transition sequences,
based upon an assumption that the transitions occur when a
Stokes line, whose dominant solution is occupied, is crossed.
Second, we have indicated a general approach to interpreting all of the
Stokes lines in a multi-level system, by using a projection technique to
selectively remove the most divergent contributions from the renormalized
Hamiltonian. And finally, we have verified the rule by means of extensive
numerical experiments.

The results of sections 3 and 4 support the rule suggested in section
2, but they do not constitute a proof. Further work must
be done to establish how the high order renormalized Hamiltonian
can be used to calculate transition probabilities directly
in the case of multi-level systems.  This might involve studies of how
the transition probabilities could be determined by applying
time-dependent perturbation theory to the high order renormalized
Hamiltonians as $t$ increases along the real axis.  This has
been successfully applied by Berry [17] for the case of two level
systems.  An alternative approach would be to investigate analytic
continuations of the adiabatic solutions of the renormalized
Hamiltonian away from the real axis, as far as the branch points. 

There is a theoretical difficulty that must be resolved
concerning the interpretation of the Stokes lines. According
to the interpretation discussed in section 4, a transition from
level $n_0$ to $n_1$ occurs on crossing a Stokes line $S({n_0,n_1})$,
where the matrix elements $H^{(k)}_{n_0,n_1}(t)$ are greatest for
real $t$. A subsequent transition from state $n_1$ to $n_2$ could then
occur if the Stokes line $S({n_1,n_2})$ were crossed at a later time.
Presumably the allowed transition sequence would then be determined by
the order in which the Stokes lines cross the real axis.  However,
the path in the complex time plane along which the Schr\" odinger
equation is integrated can be deformed away from the real $t$ axis.
A problem may arise if the two Stokes lines cross: a transition
 allowed along one path would then be forbidden along an equally
valid path. A rule based on the order in which Stokes lines cross the
real axis may therefore predict different transition
sequences from the rule in section 2.
We have not yet found a totally satisfactory
resolution of this problem. We can, however, remark that
in our investigations we found no
examples where the predictions of the two possible
rules were different. It is not clear whether it is
impossible to find an example in which the predictions differ, or
whether
such cases occur with very low probability within
our ensemble.

%
%
\section{Acknowledgements}

This work was supported by a research grant, reference GR/L02302
from the EPSRC (UK), and by the program \lq Dynamics of Complex
Systems', at the Max Planck Institute for the Physics of Complex Systems,
Dresden. We would like to thank the Nuclear Theory
Group at the University of Washington, Seattle and the Max Planck Institute 
for their hospitality during our visits.

%
%

%
%
\figure{Figure 1. The exponent characterising a non-adiabatic transition in a
two-level system is obtained by integrating the energy along a path
enclosing a branch point in the complex $t$ plane where the levels
become degenerate.}

\figure{Figure 2. (a) Energy levels as a function of real 
time for a three level
system.  The regions where the curves approach each other closely are called
\lq avoided crossings'. (b) The avoided crossings correspond to branch points
close to the real axis.}

\figure{Figure 3. (a) The sign of an eigenvector is reversed
upon making two circuits about a branch point. (b) Illustrates
this change of sign for the Landau-Zener Hamiltonian by considering
the eigenvector transported smoothly around a circuit taken to infinity
in the upper-half plane.}

\figure{Figure 4. Stokes lines $({\rm S}n)$, anti-Stokes 
lines $({\rm A}n)$, and branch cut (BC) associated with a 
single branch point.}

\figure{Figure 5. Stokes and anti-Stokes lines associated 
with a pair of branch points between different pairs of 
levels. The boundary condition is that only level $n_0$ is 
occupied as $t\to -\infty$. Case (a) allows a transition
to level $n_2$ as $t\to \infty$. Case (b) does not allow such 
a transition.}

\figure{Figure 6. Energy levels for example I of the model Hamiltonian
introduced in section 5.}

\figure{Figure 7. Energy levels for example II of the model Hamiltonian
introduced in section 5.}

\figure{Figure 8. Branch points and Stokes lines intersecting the real axis
for example I. The pairs of integers indicate which energy levels
become degenerate at each  of the branch points.}

\figure{Figure 9. Branch points and Stokes lines intersecting the 
real axis for example II.}

\table{Table 1. Branch point data for example I.}

\table{Table 2. Branch point data for example II.}

\table{Table 3. Theoretical transition sequences and actions compared
with empirical values for example I.}

\table{Table 4. Theoretical transition sequences and actions compared
with empirical values for example II.}

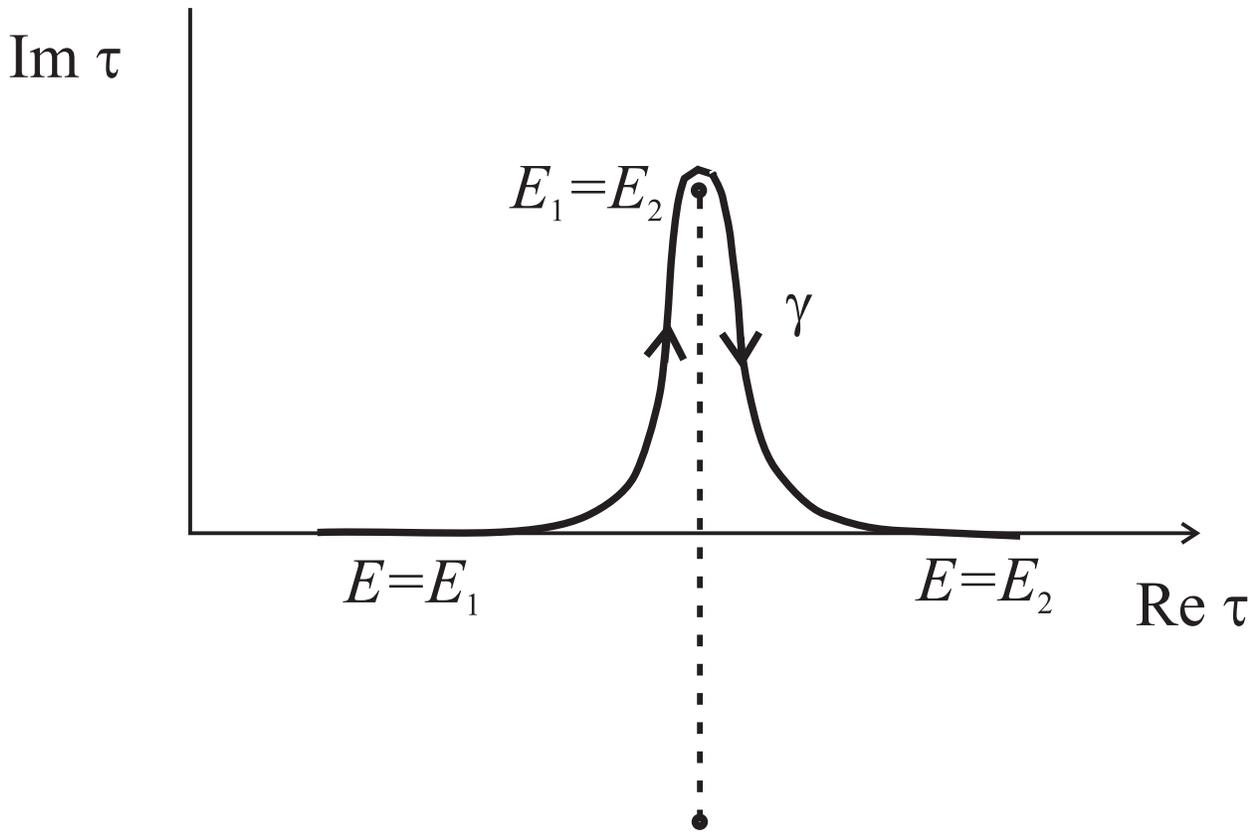
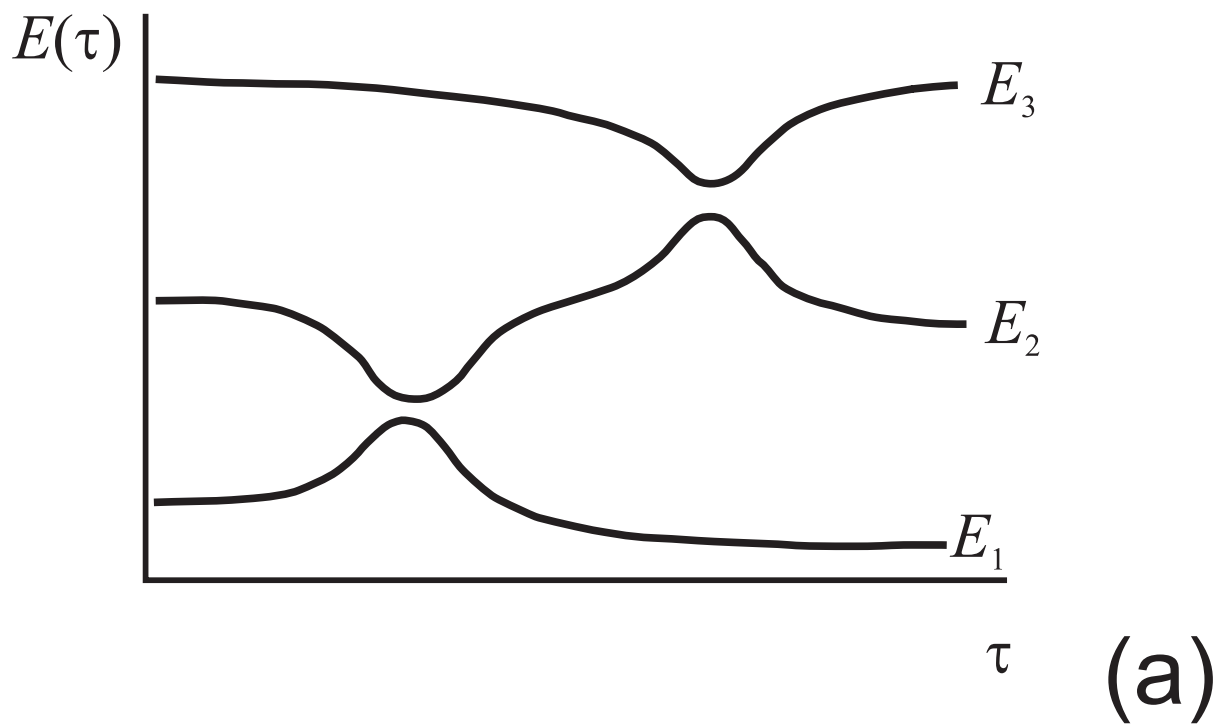
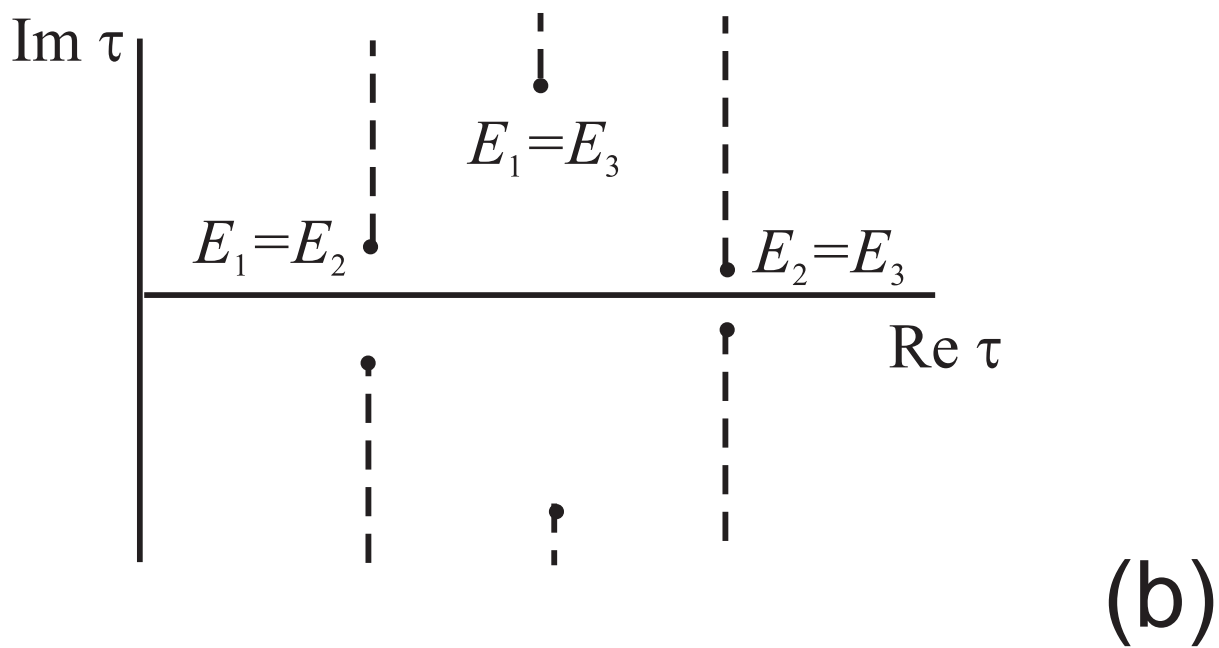
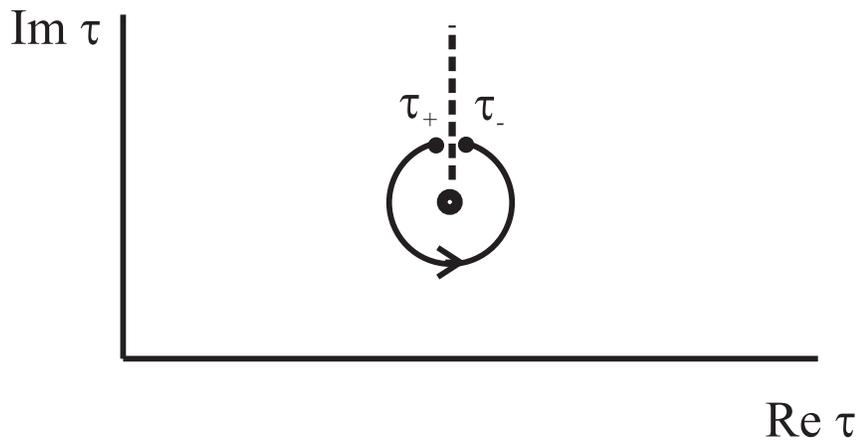
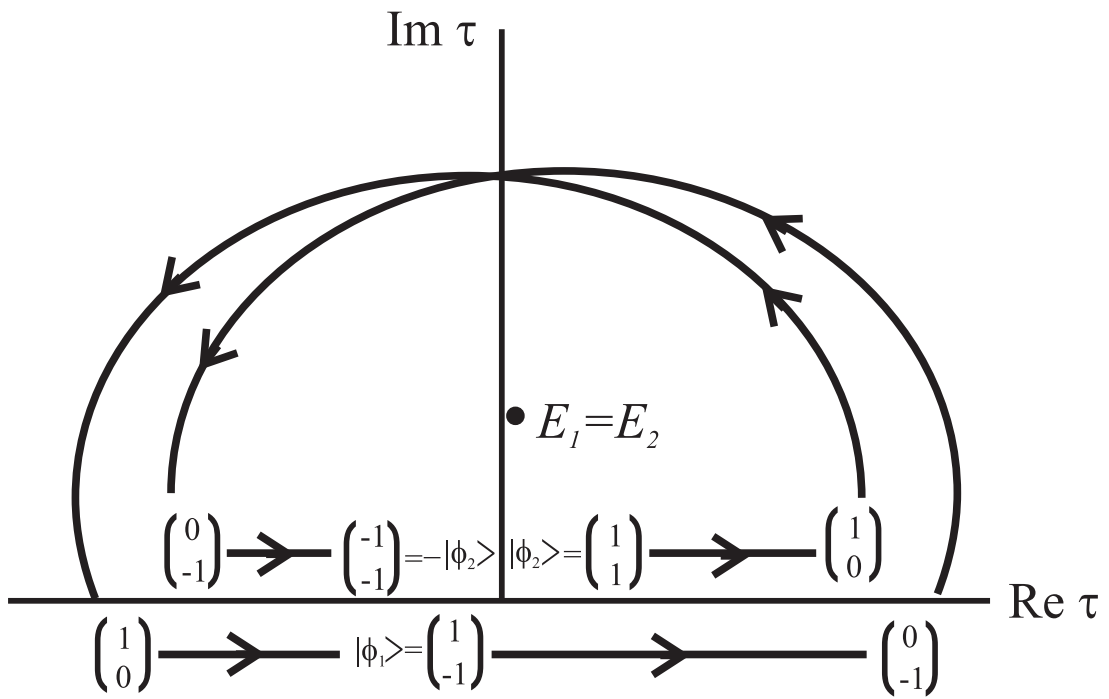
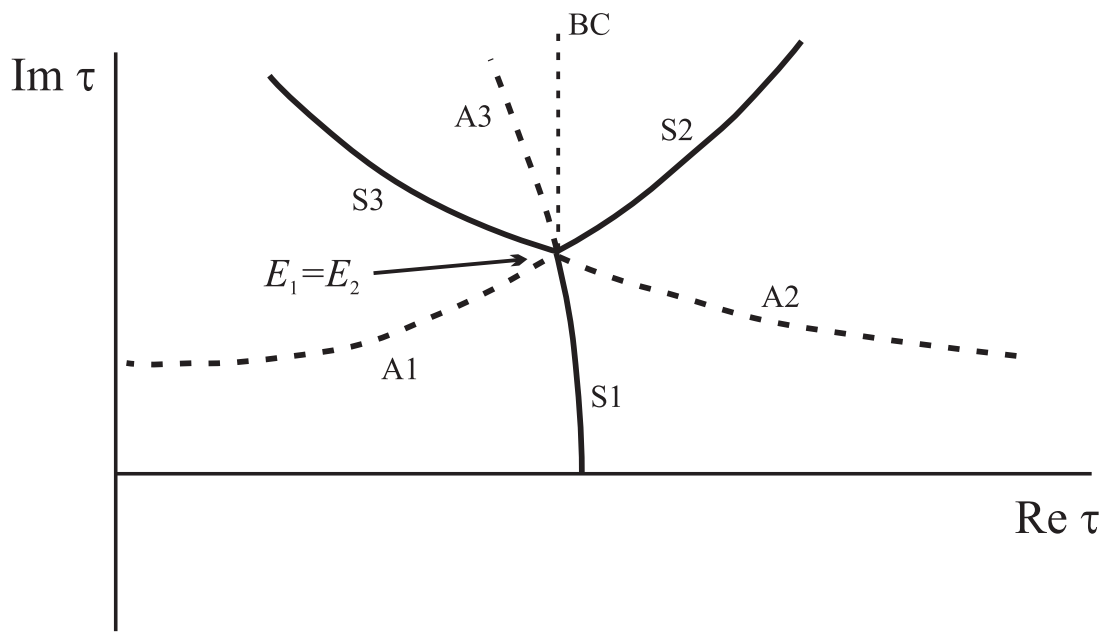
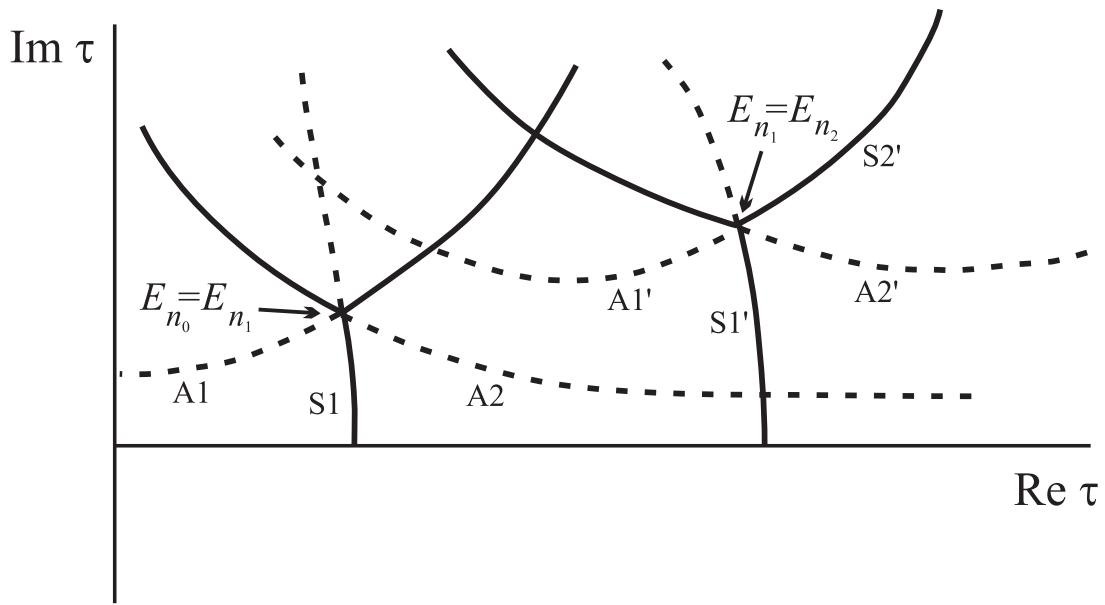
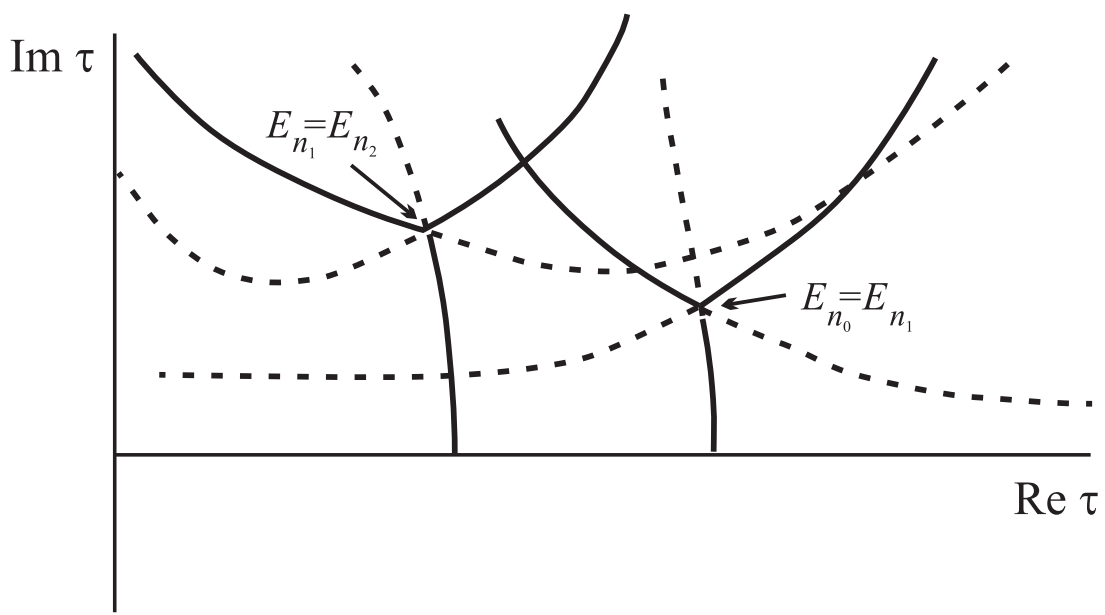
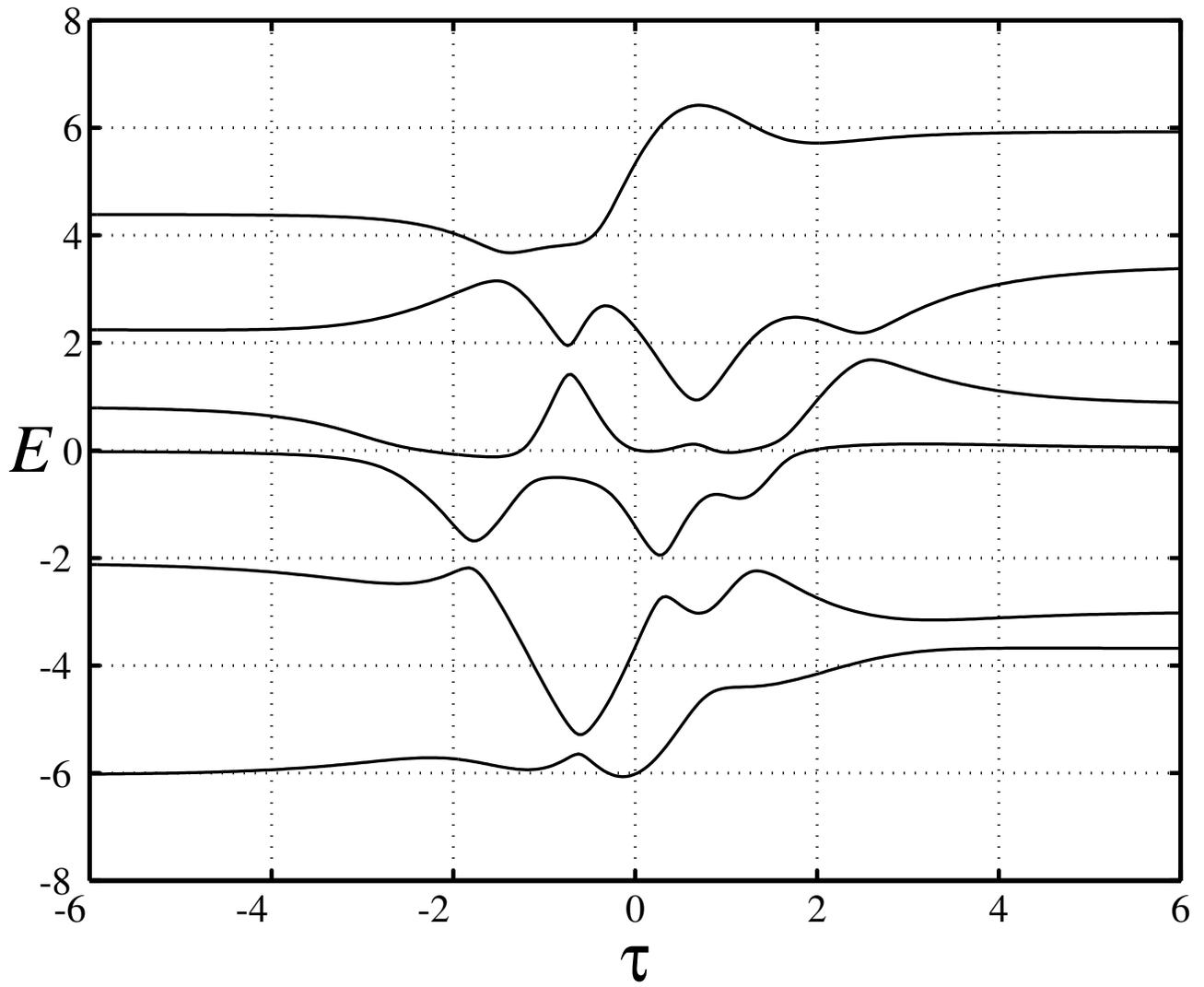
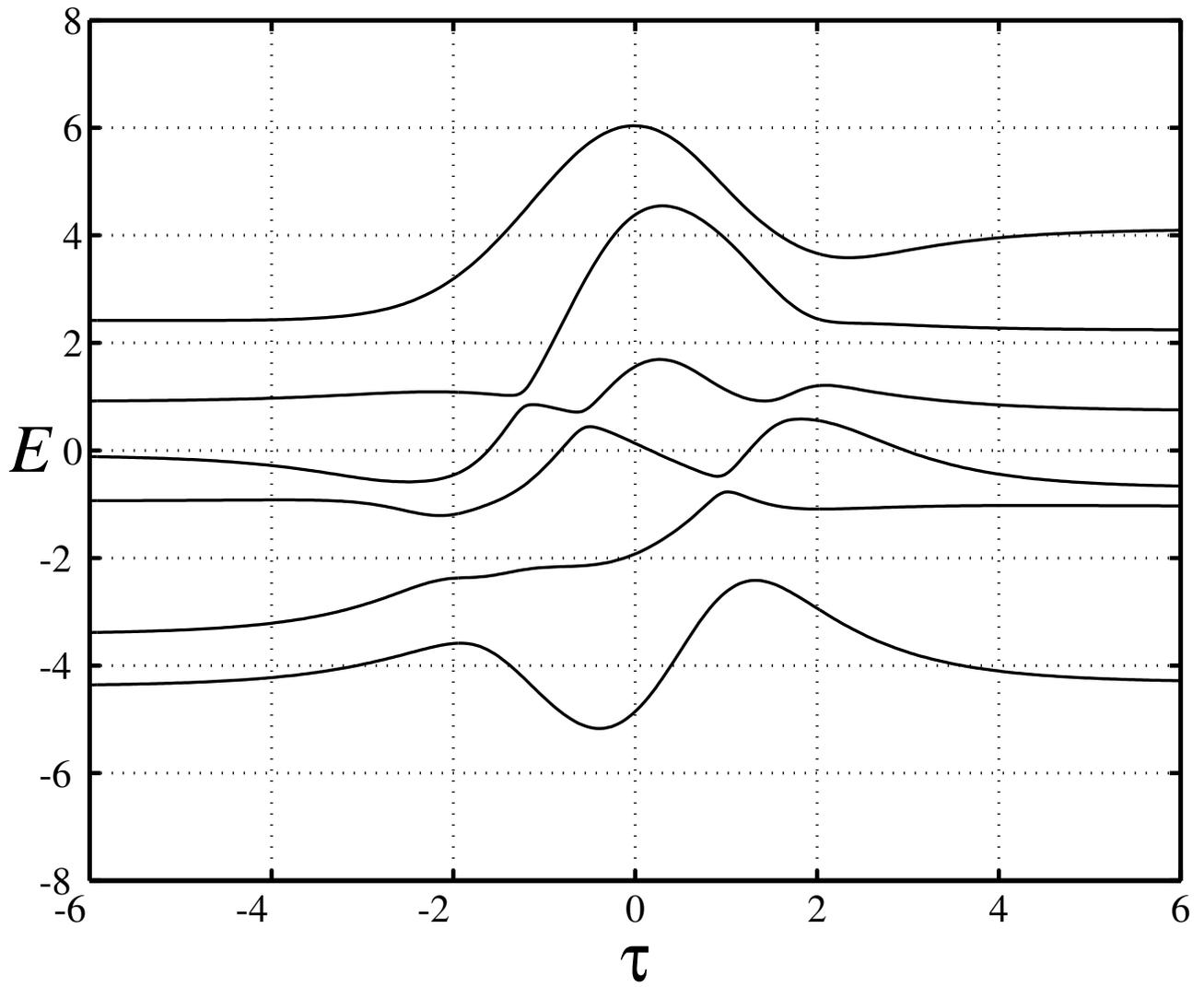
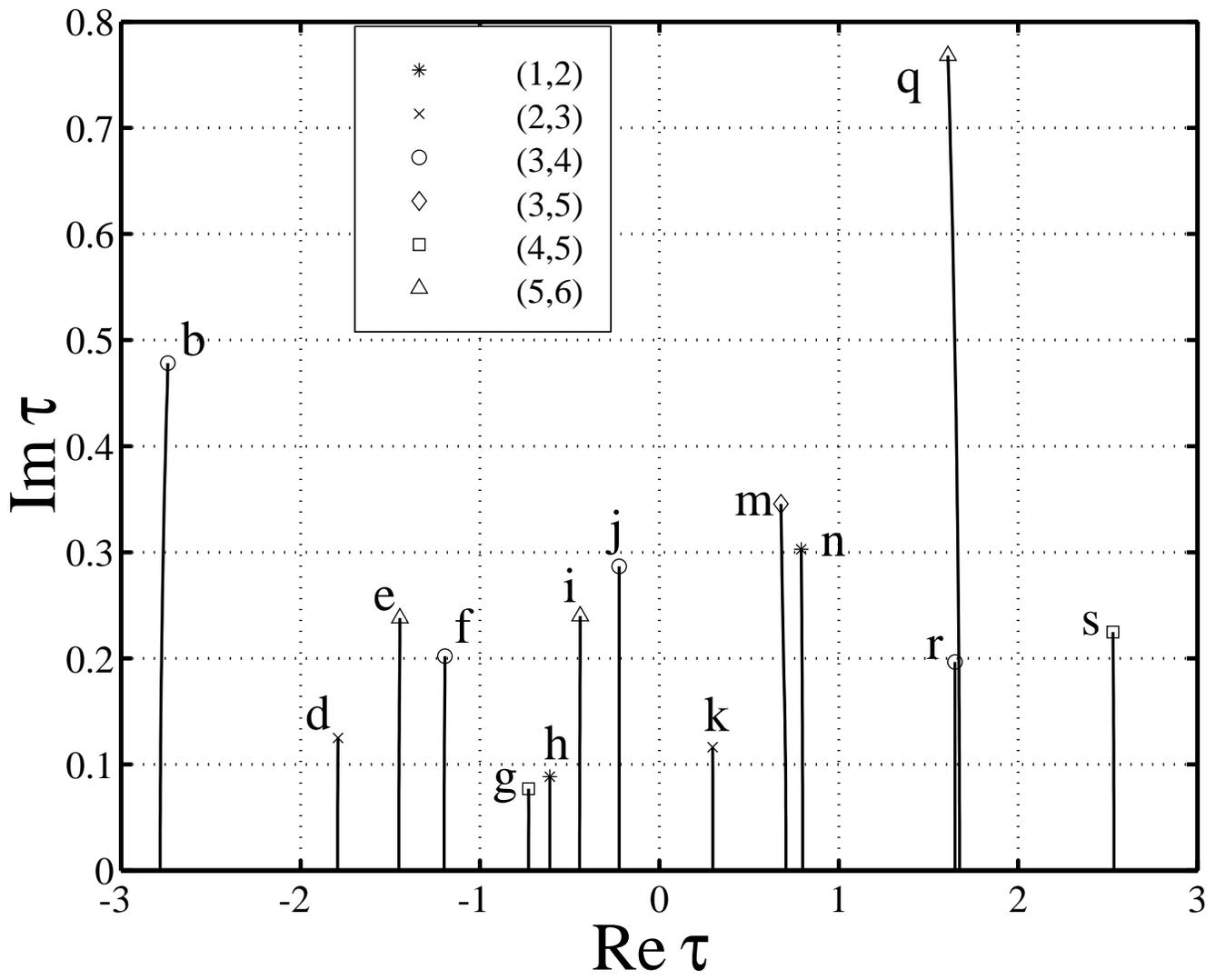
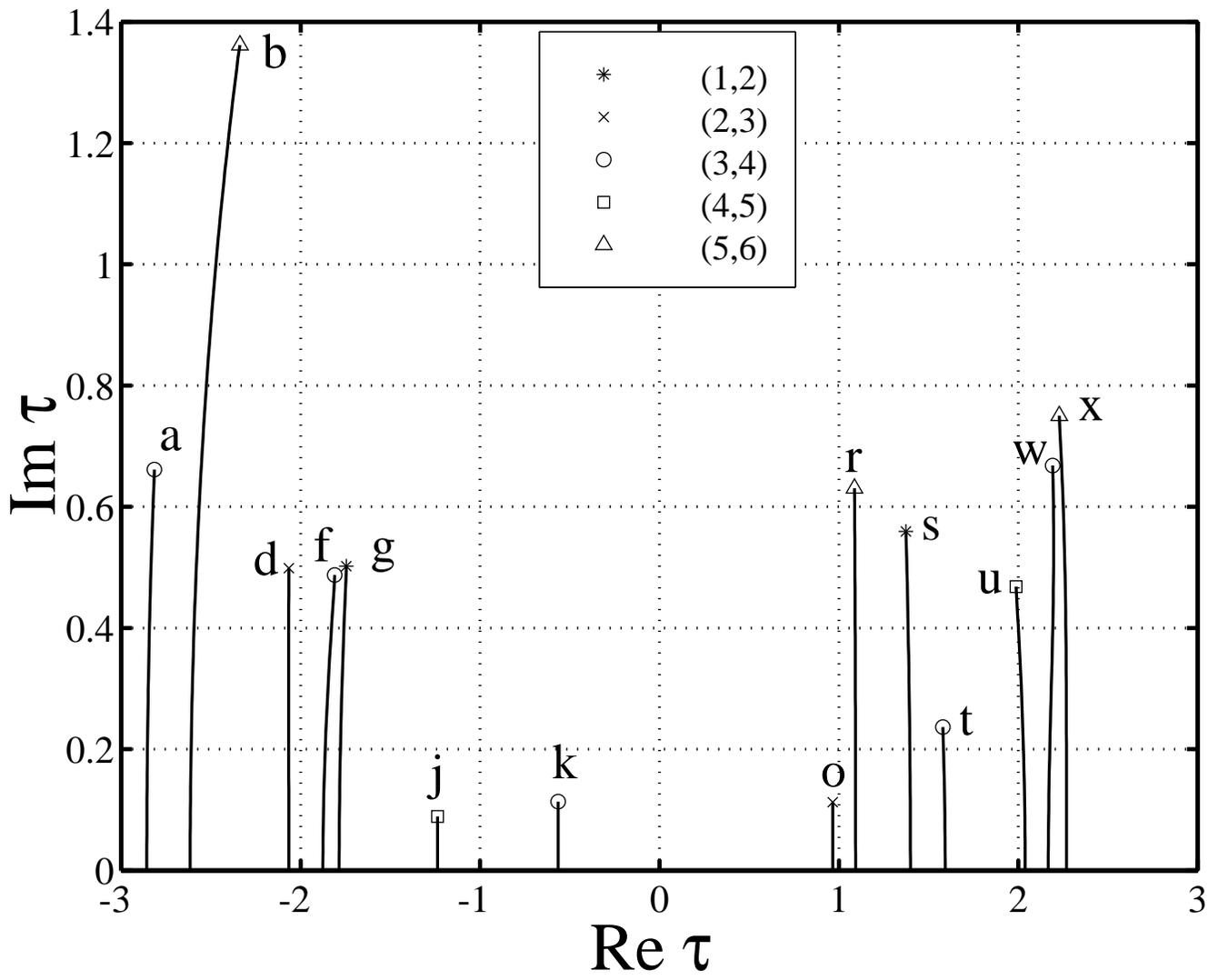
\end{document}